\documentclass[%
 aip,
 jmp,%
 amsmath,amssymb,
reprint,%
onecolumn]{revtex4-1}

\usepackage{graphicx}
\usepackage{dcolumn}
\usepackage{bm}
\usepackage{hyperref}
\usepackage{natbib}
\usepackage{amssymb}
\usepackage{amsmath}
\usepackage{stackengine}
\usepackage{pgfplots}
\usepackage{adjustbox}
\usepackage{makecell, multirow, tabularx}

\usepackage{float}
\usepackage{caption}
\usepackage{subcaption}
\pgfplotsset{compat=1.18}

\begin{document}

\title{Intraday order transition dynamics in high, medium, and low market cap stocks: A Markov chain approach}

\author{Salam Rabindrajit Luwang}
\email{salamrabindrajit@gmail.com}
\affiliation{Department of Physics, National Institute of Technology Sikkim, India-737139.}

\author{Anish Rai}
\email{anishrai412@gmail.com}
\affiliation{AlgoLabs, Chennai Mathematical Institute, Kelambakkam, India-603103.}
\affiliation{Department of Physics, National Institute of Technology Sikkim, India-737139.}

\author{Md. Nurujjaman}
\email{md.nurujjaman@nitsikkim.ac.in}
\affiliation{Department of Physics, National Institute of Technology Sikkim, India-737139.}

\author{Filippo Petroni}
\email{fpetroni@luiss.it}
\affiliation{Department of Economics, University of Chieti-Pescara, Italy-65127}
	
\date{\today}

\begin{abstract}

An empirical stochastic analysis of high-frequency, tick-by-tick order data of NASDAQ100 listed stocks is conducted using a first-order discrete-time Markov chain model to explore intraday order transition dynamics. This analysis focuses on three market cap categories: High, Medium, and Low. Time-homogeneous transition probability matrices are estimated and compared across time-zones and market cap categories, and we found that limit orders exhibit higher degree of inertia (DoI), i.e., the probability of placing consecutive limit order is higher, during the opening hour. However, in the subsequent hour, the DoI of limit order decreases, while that of market order increases. Limit order adjustments via additions and deletions of limit orders increases significantly after the opening hour. All the order transitions then stabilize during mid-hours. As the closing hour approaches, consecutive order executions surge, with decreased placement of buy and sell limit orders following sell and buy executions, respectively. In terms of the differences in order transitions between stocks of different market cap, DoI of orders is stronger in high and medium market cap stocks. On the other hand, lower market cap stocks show a higher probability of limit order modifications and greater likelihood of submitting sell/buy limit orders after buy/sell executions. Further, order transitions are clustered across all stocks, except during opening and closing hours. The findings of this study may be useful in understanding intraday order placement dynamics across stocks of varying market cap, thus aiding market participants in making informed order placements at different times of trading hour.

\end{abstract}

\maketitle


\section{Introduction}
\label{sec:Intro}

The choice of order types and submission strategies are crucial for trading in stock markets~\cite{peterson2002order,harris1996market,cho2000probability}. Interestingly, the choice of order type is not considered to be a random decision and, therefore, a potential selectivity bias exists~\cite{peterson2002order}. Traders, who prioritize speed over price and seek immediate execution at the current best listed price, submit market orders. With limit orders, traders have the possibility to improve the price of execution but face the risk of non-execution~\cite{hollifield2004empirical,harris1996market,lo2003order}. As a result, limit orders are frequently modified, partially canceled, or fully withdrawn prior to execution to mitigate adverse selection risks~\cite{foucault1999order,harris1998optimal,large2004cancellation}. Given this background, it may be beneficial for a trader to get an idea of the likelihood for a particular order to transition to other order types as they may use this information to optimize their order-submission strategy. For example, what is the probability that the next transaction will be an order deletion following a limit order submission or an order execution? On the other hand, it is well known that stock market activity varies depending on the time of day, with distinct order types being prevalent in different times of a trading day\cite{garvey2009intraday,admati1988theory}. Trading frequency is known to be higher just after the start and just before the end of the market~\cite{toyabe2022stochastic}. The dynamics of the conditional probability of order execution is also different for different times of a day~\cite{ma2004dynamic}. Therefore, it is essential to analyze how the probabilities of transitioning between different stock market order types vary throughout the trading day. By doing so, traders can better anticipate market behavior and adjust their order placement strategies according to the specific trading hour.

Order placement strategies and the overall trading process also varies depending on the market cap of stocks. Higher market cap stocks are generally traded more with higher limit order activities and shorter duration between order events~\cite{lo2015resiliency}. The rate of order recovery is also high, showing a consistent high level of resiliency. However, greater variation in resiliency is observed across lower market cap stocks~\cite{lo2015resiliency}. These stocks also tend to have lower order arrival rates and limit order traders, facing longer waiting times. As a result, such less frequently traded stocks have greater variability in order flow~\cite{easley1996liquidity}. Additionally, information asymmetries are more significant for smaller firms~\cite{hasbrouck1991measuring} and hence, adverse selection effects are more severe~\cite{frey2008liquidity}. Market cap also has a direct impact on liquidity: Less liquid stocks often belong to the lower market cap segments~\cite{lo2015resiliency}. The transaction costs for smaller market cap stocks are likely larger than higher market cap stocks~\cite{keim1997transactions}. Recognizing these distinct variations, it is crucial to analyze the order transitions between stocks of different market cap at different times of a trading day. Such analysis may provide insights into how different market cap stocks react to market conditions according to the time of the day, thus helping market participants improve trading strategies by making informed order placements. Hence, in this paper, we study the intraday order transition dynamics for high, medium and low market cap stocks.

Various studies have been carried out to study the intraday variations and order placement dynamics.
Intraday patterns have been observed for returns~\cite{wood1985investigation}, trading volume~\cite{chung1999limit,kluger2011intraday}, number of trades and the number of shares per trade~\cite{ho1991behaviour}, liquidity~\cite{brockman1999analysis} and bid-ask spreads~\cite{lee1993spreads,chan1995market}. Several studies have also tried to explain these different intraday patterns~\cite{admati1988theory,brock1992periodic,chung1999limit,garvey2007intraday}. Studies on order placement dynamics~\cite{ellul2007order, biais1995empirical,griffiths1998information} have found that there is a positive serial correlation in order flow, which suggests that follow-on order strategies dominate. However, when private information is considered, it was found that order type is serially uncorrelated~\cite{kaniel2006so}. Further, limit orders submitted at midday took significantly longer time to execute~\cite{garvey2007intraday}. Intraday studies in financial markets have also been carried out using high-frequency data to study spread, where there is a high
rate of order cancellation when liquidity demands are high~\cite{ammar2020high}. Additionally, the market open and close serve as specific clustering points~\cite{admati1988theory} and the demand for transactions during these intervals is higher than during the middle of the trading day~\cite{brock1992periodic}. Despite all these studies, to the best of our knowledge, no study have been carried out to analyze the order transition dynamics at different times of a trading day for high, medium and low market cap stocks.

In order to study the order transition dynamics, we have employed high-frequency tick-by-tick order transaction data of stocks listed in NASDAQ100, as shown in Table~\ref{tab:DataSample}. Beginning from the opening ($9:30:00.000$) to the closing ($16:00:00.000$) hour, market participants place different types of orders (in total, ten, for this study), as shown in Table~\ref{tab:EventDescription}. Thus, we have a sequence of orders consisting of ten types for the duration from the opening to the closing hour. Analyzing this kind of categorical data sequence require different approaches, among which Markov-based models are commonly used~\cite{rabindrajit2024high,van2018simple,fitzpatrick2020thermodynamics}. In the field of finance and stock market, Markov-based models have been employed to investigate various phenomena~\cite{cont2012order,shiyun1999new,d2011semi}. Particularly, in our previous work, a discrete-time Markov chain model was developed to understand order transition dynamics of stocks belonging to six different sectors during the US–China trade war of 2018~\cite{rabindrajit2024high}. Intraday dynamics of the Nikkei index futures prices, trading volumes, and spreads were studied by using the methodology of Markov chains~\cite{shiyun1999new}. Markovian jump-diffusion process have been also utilized to study the intraday dynamics of the limit order book, assuming that the frequency of order arrivals is large~\cite{cont2012order}. A study on high frequency price dynamics was also carried out under the assumption that the intraday returns are described by a discrete time homogeneous semi-Markov which depends on a memory index. The index took into account high and low volatility periods~\cite{d2011semi}. Despite these studies, to the best of our knowledge, no study has explored the intraday order transition dynamics for different market cap stocks using Markov-based models. 

In the present study, we use a first-order time-homogeneous discrete time Markov chain (DTMC) model to examine the intraday order transition dynamics. The choice of a first-order DTMC is motivated by its relative simplicity, clarity of interpretation and minimal parameter estimation requirements, while still reliably calculating important measures such as steady-state distribution~\cite{rabindrajit2024high}.
The model assumes that the probability of transitioning to a future state depends exclusively on the current state, independent of prior history (first-order Markov property), and that transition probabilities remain constant over time (time homogeneity~\cite{masseran2015markov}). We analyze and compare these time-homogeneous transition probability matrices (TPMs) to capture order dynamics at distinct trading periods for stocks grouped by market cap. Additionally, we evaluate stationary distributions to assess long-term order behavior and compare these distributions with Jensen-Shannon divergence. Finally, we perform cluster analysis on TPMs to identify patterns in order transitions specific to different time intervals. The findings of this study not only complement prior research~\cite{ellul2007order, biais1995empirical,griffiths1998information,roth1988deadline,harris1998optimal,hollifield2004empirical} but also uncover new insights into the order transition dynamics at different times of the trading day for high, medium, and low market cap stocks. 

The rest of the paper is organized as follows: Sec.~\ref{sec:DD} describes the high-frequency stock market tick-by-tick order transaction data, Sec.~\ref{sec:Method} describes the methods used for the analysis. The results are discussed in Sec.~\ref{sec:R}. Finally, Sec.~\ref{sec:Conc} gives the conclusion of our study.

\section{Data Description}
\label{sec:DD}

The accessibility of high-frequency stock market data at a micro level, which includes real-time recordings of trades such as the order type, as well as the corresponding prices, volumes, and time stamps, opened new perspectives for the empirical analysis of the market microstructure of financial markets. For this study, Algoseek~\cite{Algo} provided the high-frequency tick-by-tick order transaction data. The provided full data format contains information of all the types of orders placed starting from 04:00:00 EST to 20:00:00 EST for all the listed stocks in NASDAQ100. The length of this data varies daily, typically ranging in the tens of millions, with data sizes between $20$ to $40$ GB. processing~\cite{rabindrajit2024high}. The sample structure of the full data format is provided in Table~\ref{tab:DataSample}. It contains eight columns, with the first column showing the date, the second column showing the timestamp, the third column showing the order ID, the fourth column showing the event type, the fifth column showing the ticker symbol of the traded stock, the sixth column showing the price at which the transaction occurred, the seventh column showing the number of stocks traded, and the eighth column showing the exchange on which the trade took place. EmEditor, a text editor, was used to extract and pre-process the data for particular stocks choosen for this study.

\begin{table}[H]
\centering
\caption{Table shows a sample dataset of the high-frequency tick-by-tick order data of stocks listed in NASDAQ100.}
\label{tab:DataSample}
\begin{tabular}{|c|c|c|c|c|c|c|l|} 
\hline 
\textbf{Date} & \textbf{Timestamp} & \textbf{Order Id.} & \textbf{Event Type} & \textbf{Ticker} & \textbf{Price} & \textbf{Quantity} & \textbf{Exchange} \\
\hline 
2018-11-06 & 4:00:00.002 & 12011 & ADD-BID & AAPL & 164.99 & 100 & NASDAQ \\
\hline 
2018-11-06 & 4:00:00.032 & 12056 & ADD-ASK & AAPL & 194.99 & 500 & NASDAQ \\
\hline
2018-11-06 & 4:00:00.112 & 13473 & ADD-BID & XLF  &  67.50 & 300 & NASDAQ \\
\hline
... & ... & ... & ... & ... & ... & ... & ... \\ 
\hline
2018-11-06 & 9:30:00.156 & 89017 & DELETE-BID & GOOGL & 0 & 100 & NASDAQ \\ 
\hline
2018-11-06 & 9:30:01.006 & 83907 & ADD-BID & INTC & 123.70 & 200 & NASDAQ \\
\hline
... & ... & ... & ... & ... & ... & ... & ... \\ 
\hline
2018-11-06 & 16:00:00.000 & 123483 & DELETE-BID & AMD & 0 & 150 & NASDAQ \\ 
\hline
... & ... & ... & ... & ... & ... & ... & ... \\ 
\hline
2018-11-06 & 20:00:00.000 & 547324 & DELETE-ASK & NVDA & 0 & 40 & NASDAQ \\
\hline
\end{tabular}
\end{table}

The event type in the fourth column of Table~\ref{tab:DataSample} are the different types of orders placed from 04:00:00 EST to 20:00:00 EST. The order types (in total, ten), their abbreviations (for the analysis in this study) and the definitions are shown in Table~\ref{tab:EventDescription}. Considering these order types as states of the Markov chain, we will analyze the transition probabilities between them in this study. Detailed descriptions of the Markov chain and other methods used for the analysis are provided in the subsequent Sec.~\ref{sec:Method}.

\begin{table}[H]
\centering
\caption{Names and descriptions of the type of orders found in the dataset.}
\label{tab:EventDescription}
\begin{tabular}{|l|c|p{6.5cm}|}
\hline 
\textbf{Order (State)} & \textbf{Abbreviation} & \textbf{Description} \\
\hline 
ADD-BID      & AB & Add Bid order \\
\hline 
ADD-ASK      & AA & Add Ask order \\
\hline
DELETE-BID   & DB & Delete outstanding Bid order in full \\
\hline
DELETE-ASK   & DA & Delete outstanding Ask order in full \\
\hline
FILL-BID     & FB & Execute outstanding Bid order in full \\
\hline
FILL-ASK     & FA & Execute outstanding Ask order in full \\
\hline
EXECUTE-BID  & EB & Execute outstanding Bid order in part \\
\hline
EXECUTE-ASK  & EA & Execute outstanding Ask order in part \\
\hline
CANCEL-BID   & CB & Cancel outstanding Bid order in part \\
\hline
CANCEL-ASK   & CA & Cancel outstanding Ask order in part \\
\hline
\end{tabular}
\end{table}

\section{Methodology}
\label{sec:Method}
In this section, we discussed the methodologies that are used for the analysis of stock market intraday order transition dynamics using high-frequency tick-by-tick order transaction data.

\subsection{G-test of independence}
\label{subsec:G-test}
To examine whether the sequence of high-frequency stock market order display memory, we employed the G-test of independence~\cite{berrett2021usp,ahad2019applicability,buancioiu2021accelerating}, which is a statistical hypothesis test that evaluates the likelihood of the null or alternative hypothesis being true, based on observed empirical data. The null and alternative hypotheses are:

\begin{equation*}
\begin{aligned}
      & H_0: \text{Observed data are independent}\\
      & H_1: \text{Observed data are not independent}
  \end{aligned}    
\end{equation*}

The likelihood of $H_0$ or $H_1$ is quantified by the G-statistic, which is defined as~\cite{ahad2019applicability,buancioiu2021accelerating}

\begin{equation}
    \text{G} = 2 \sum_{i, j} O_{ij} \log\left(\frac{O_{ij}}{E_{ij}}\right),
    \label{eqn:G}
\end{equation}

where $O_{ij}$ are the observed frequencies and $E_{ij}$ are the corresponding expected frequencies, computed as~\cite{berrett2021usp}

\begin{equation}
    E_{ij}=\frac{(\sum_{i=1}^{n}O_{ij})(\sum_{j=1}^{n}O_{ij})}{(\sum_{i=1}^{n}\sum_{j=1}^{n}O_{ij})},
\end{equation}
 where $n$ is the number of observations. The G-test statistic is distributed as $\chi^2$
with degree of freedom, $d_f = (r-1)(c-1)$, where $r$ and $c$ are the number of rows and columns of the contingency table of the observed data~\cite{berrett2021usp}.

The G-test statistic determines how likely the observed data are independent by evaluating the probability of obtaining the value of G as given by Eq.~\ref{eqn:G} under its limiting probability distribution, $\chi^2$~\cite{ahad2019applicability,buancioiu2021accelerating}. This results in a p-value which is the conditional probability of observations as extreme than the results indicated by the G statistic, assuming the null hypothesis, $H_0$ holds true.
The lower the p-value, the less probable the observations are under $H_0$; therefore, for low p-values, the $H_0$ may be rejected, and $H_1$ may be considered as true instead~\cite{berrett2021usp,ahad2019applicability,buancioiu2021accelerating}. The significance level of the test for p-values under which $H_0$ is set at $5\%$ for the study. After validating $H_1$ as the true hypothesis, we focus exclusively on the dependency between two consecutive variables in the data sequence in this study. This enables the application of the Markov property, which is discussed in the subsequent subsection.

\subsection{Markov Chain}
Markov chain (MC) belongs to a category of stochastic processes that are highly effective in describing sequence of categorical events~\cite{rabindrajit2024high,de2019markov}. Such stochastic processes can be broadly categorized into four types: discrete-time MC with discrete states, discrete-time MC with continuous states, continuous-time MC with discrete states, and continuous-time MC with continuous states~\cite{gao2020markov}. For this study, we consider discrete-time MC with discrete states, as described below.

\subsubsection{Discrete-Time Markov Chain (DTMC)}
A DTMC refers to a series of random variables denoted as $X_1, X_2, \ldots, X_n$, which follows the Markov property~\cite{spedicato2016markovchain}, where the transition probabilities for future states depend only on the current state~\cite{shamshad2005first,tang2020markov}. In the context of a first-order DTMC, the future state, denoted as $X_{n+1}$ relies exclusively on the current state, represented as $X_n$~\cite{spedicato2016markovchain}. The set of possible states is $S = \{s_1,s_2,....,s_r\}$, where $r=10$ for the ten order types. The probability to move from state $s_i$ to state $s_j$ in one step (or transition) is called transition probability, denoted by $p_{ij}$~\cite{rabindrajit2024high} as,
\begin{equation}
    p_{ij} = P(X_{n+1} = s_j\mid X_n = s_i)
\end{equation}

The probability distribution of the transitions can be represented by
a transition probability matrix, $P = {(p_{ij})}_{i,j}$, where each element of position $(i, j)$ represents the transition probability $p_{ij}$. For $r$ states, $P$ has the form,~\cite{rabindrajit2024high,shamshad2005first,tang2020markov}
\begin{equation}
\label{eqn:6}
P = 
\begin{pmatrix}
p_{11} & p_{12} & \cdots & p_{1r} \\
p_{21} & p_{22} & \cdots & p_{2r} \\
\vdots  & \vdots  & \ddots & \vdots  \\
p_{r1} & p_{r2} & \cdots & p_{rr} 
\end{pmatrix}
\end{equation}
subject to 
\begin{align}
    & 0\leq p_{ij}\leq 1, \forall i,j \in S, \\
    & \sum_{j=1}^{r} p_{ij} = 1, \forall i \in S.
\end{align}

The Maximum Likelihood Estimation method~\cite{anderson1957statistical} was employed to estimate $p_{ij}$. By utilizing this approach, the number of transitions from state $i$ to state $j$ in the data sequence represented as $n_{ij}$, was used to derive the maximum likelihood estimate of the one-step transition probability~\cite{shamshad2005first,masseran2015markov}, given as $\displaystyle p_{ij} = \frac{n_{ij}}{\sum_{j=1}^{r} n_{ij}}$. A modified MATLAB code is used for $p_{ij}$ estimation and the original version of the code can be found in Ref.~\cite{JesseMATLABcode}. We utilize these $p_{ij}$ values to analyze the intraday order transition dynamics. They are further used for estimating the stationary distribution of each order type, as detailed below. 

\subsubsection{Stationary Distribution}

The stationary distribution of a Markov chain describes the long-term probabilities of being in each state, which remain constant over time~\cite{holmes2021discrete}. For a finite and ergodic (i.e., aperiodic and irreducible) Markov chain, the transition probabilities stabilize to this distribution as the number of steps increases indefinitely~\cite{holmes2021discrete}. Let \(\pi_j\) represent the probability of being in state \(j\) in the long run. For an ergodic chain, the stationary distribution is unique and it satisfies the following equations.~\cite{holmes2021discrete,rabindrajit2024high}:
\begin{align}
    \pi_j &= \sum_{i=1}^{r} \pi_i p_{ij} \quad \text{(Balance equation)}, \label{eq:balance} \\
    \sum_{j=1}^{r} \pi_j &= 1 \quad \text{(Probability normalization)}. \label{eq:norm}
\end{align}

PyDTMC, a Python package~\cite{belluzzo2024pydtmc} is employed for determining stationary distributions. To compare the similarity between these probability distributions across different time-zones [shown in Table~\ref{tab:time_zones}], we employ a well-known divergence measure, Jensen-Shannon Divergence, as explained in the next subsection.

\subsection{Jensen-Shannon Divergence}
Let $p$ and $q$ be two probability distributions where their elements are non-negative and sum up to one. To quantify the difference between $p$ and $q$, Jensen-Shannon Divergence (JSD) measure is utilized, which is defined as follows~\cite{grosse2002analysis,nielsen2019jensen}:

\begin{equation}
    \text{JSD}\hspace{0.1cm}(p, q) = \frac{1}{2} \biggl[\text{KLD}\biggl(p, \frac{p+q}{2}\biggr) + \text{KLD} \biggl(q, \frac{p+q}{2}\biggr)\biggr],
\end{equation}
where KLD is the Kullback-Leibler divergence, defined as~\cite{nielsen2019jensen}:
\begin{equation}
    \text{KLD}\hspace{0.1cm}(u, v) = \sum_{i} u_i \log \frac{u_i}{v_i},
\end{equation}

where $u$ and $v$ are discrete probability distributions. $u_i$ and $v_i$ are the probability of the $i^{th}$ element in distribution $u$ and $v$, respectively. The logs are taken to base two. JSD is always nonnegative and symmetric, and it is zero if and only if $u = v$~\cite{mateos2017detecting}. We further aim to cluster the TPMs corresponding to various time-zones. To facilitate this clustering, we first applied Principal Component Analysis to reduce the dimensionality of the TPMs, as described in the following subsection.

\subsection{Principal Component Analysis}

Principal Component Analysis (PCA) is a technique used to
reduce dimensions of a complex dataset~\cite{liu2011integrating} while preserving as much of the original information as possible. The reduced dimensions are referred to as the principal components. The specific steps to compute the principal components are as follows ~\cite{liu2011integrating,sondergaard2024enhanced}:

\begin{enumerate}
    \item Assume the data matrix with $m$ variables, $P_1, P_2, . . . , P_m, l$ times observations.
    \begin{equation}
P = 
\begin{pmatrix}
P_{11} & P_{12} & \cdots & P_{1m} \\
P_{21} & P_{22} & \cdots & P_{2m} \\
\vdots  & \vdots  & \ddots & \vdots  \\
P_{l1} & P_{l2} & \cdots & P_{lm} 
\end{pmatrix}
\end{equation}
\item  Normalize the original data as,
\begin{equation}
    Y_{ti} = \frac{(P_{ti}-\overline{P_t)}}{S_t},
\end{equation}
where $\overline{P_t} = \displaystyle \frac{1}{l}\sum_{t=1}^{l} P_{ti}$ and $S_t = \displaystyle \sqrt{\frac{1}{l-1} \sum_{i=1}^{l} (P_{ti} - \overline{P}_t)^2}$ are mean and standard deviation. For convenience, the normalized $Y_{ti}$ is still denoted as $P_{ti}$. 

\item Let $\lambda_1\geq\lambda_2\geq.....\geq \lambda_m \geq 0$ and $\alpha_1, \alpha_2,.....,\alpha_m$ be the eigenvalues and the corresponding eigenvectors
of covariance matrix of normalized data, $P_{ti}$.
The eigenvectors represent the directions of the axes where there is the most variance (i.e., the principal components), and the eigenvalues represent the magnitude of the variance along these directions.

\item The $i^{th}$ principal component is such that $F_i=\alpha_{i}^T P $, where $i=1,2,.....,m$. Generally, $\displaystyle\frac{\lambda_k}{\sum_{i=1}^{m} \lambda_i}$ is called the contribution rate of the $k^{th}$ principal component and
$\displaystyle\sum_{i=1}^{k} \frac{\lambda_i}{\sum_{i=1}^{m} \lambda_i}$ is called the cumulative contribution rate of the first k principal components.

\item If the cumulative contribution rate exceeds 80\%, the first $k$ principal components contain the most information of m original variables.

\end{enumerate}

In our analysis, we have three cases of $P$, i.e., for high, medium, and low market cap. Each matrix is \( 6 \times 100 \) in dimension, with the first row representing the TPM of T1 with dimension of \( 1 \times 100 \), the second row corresponding to T2 with dimension of \( 1 \times 100 \), and so on, up to T6 in the last row (i.e., $l = 6, m = 100$). Using PCA, the three \( 6 \times 100 \) matrices are reduced to \( 6 \times 2 \) matrices by retaining only the first two principal components. Finally, we cluster these three resulting \( 6 \times 2 \) matrices using DBSCAN in a 2-D plane.

\subsection{Density-Based Spatial Clustering of Applications with Noise}

Density-Based Spatial Clustering of Applications with Noise (DBSCAN) is a density-based clustering algorithm that clusters data points into a group. It estimates the density by counting the number of points in a fixed-radius neighborhood and considers two points as connected if they lie within each other’s neighborhood~\cite{li2023urban,kazemi2017spatio}. 

\vspace{0.2cm}
To describe this clustering algorithm, the following definitions needs to be understood.~\cite{kazemi2017spatio}
\begin{enumerate}
    \item MinPts is the minimum number of points required within a point's Eps-neighborhood for that point to be classified as a core point.
    \item Eps is the radius of the neighborhood around a point, specifying the maximum distance within which other points are considered to be in the same neighborhood.
    \item A border point contains less points than MinPts within Eps, but is in the neighborhood of a core point whereas a noise point is any point that is not a core point, nor a border point.
\end{enumerate}    

The value of MinPts is given manually. To establish the value of the parameter, Eps, k-distance graph approach is utilized~\cite{kazemi2017spatio}. Finally, the steps of the DBSCAN algorithm are~\cite{kazemi2017spatio}:

\begin{enumerate}
    \item Start by marking all points as unvisited and sequentially process each one, unless it has already been processed. Establish the values of the two parameters, Eps and MinPts.
    \item Assess and label each point:
    \begin{enumerate}
        \item Label as a core point if it has at least MinPts within its Eps-neighborhood and initiate a cluster.
        \item If the point does not meet the criteria to be a core point but is reachable from any existing core point, label it as a border point.
    \end{enumerate}
    \item Repeat the process for each point, ensuring all are evaluated for clustering. Iterate over all points until every point has been visited and appropriately labeled. This ensures that all potential clusters are fully explored.
    \item At the end of the process, label all remaining unvisited points as noise, since they are not part of any cluster.
\end{enumerate}

The Python package, Scikit-learn~\cite{pedregosa2011scikit} has been employed for the PCA and DBSCAN analysis and the corresponding results are provided in Sec.~\ref{result:PCA_DBSCAN}. The following section outlines the criteria for selecting stocks, the segmentation of time-zones, and the methodology for choosing days in the analysis of intraday order transitions.  

\section{Stock Selection criteria and Time-zone division}
\label{result:Stock_selection}

The study is conducted for -- High Market Cap (HMC), Medium Market Cap (MMC), and Low Market Cap (LMC) stocks. For HMC, we identified stocks ranked $1^{st}$ to $20^{th}$ by market cap, then chose five stocks from different sectors to minimize sector bias. Similarly, five stocks were chosen from $41^{st}$ to $60^{th}$ and $81^{st}$ to $100^{th}$ for MMC and LMC, respectively. The selected stocks for HMC, MMC and LMC are shown below in Table~\ref{tab:SelectedStocks}. 

\begin{table}[H]
\centering
\caption{Selected stocks for HMC, MMC and LMC listed in NASDAQ100.}
\label{tab:SelectedStocks}
\begin{tabular}{|p{4.3cm}|p{4.3cm}|p{4.3cm}|} 
\hline 
\textbf{HMC $(1^{st}-20^{th})$} & \textbf{MMC $(40^{th}-60^{th})$} & \textbf{LMC $(80^{th}-100^{th})$} \\ 
\hline 
Amazon.com Inc[AMZN]\newline (Consumer Services) & AbbVie Inc[ABBV]\newline (Healthcare) & Broadcom Inc[AVGO]\newline (Information Technology) \\ 
\hline 
Johnson \& Johnson[JNJ]\newline (Healthcare) & HSBC Holdings plc[HSBC]\newline (Finance) & Booking Holdings Inc[BKNG]\newline (Consumer Services) \\ 
\hline
JPMorgan Chase \& Co[JPM]\newline (Finance) & Netflix Inc[NFLX]\newline (Consumer Services) & Bristol-Myers Sq Co[BMY]\newline (Healthcare) \\ 
\hline
Microsoft Corp[MSFT]\newline (Information Technology) & Oracle Corp[ORCL]\newline (Information Technology) & Nike Inc[NKE]\newline (Consumer Goods) \\ 
\hline
Exxon Mobil Corp[XOM]\newline (Oil \& Gas) & PepsiCo Inc[PEP]\newline (Consumer Goods) & Union Pacific Corp[UNP]\newline (Industrials) \\ 
\hline 
\end{tabular}
\end{table}

The trading days considered for this study are: $07-11-2018, 15-11-2018, 28-11-2018, 06-12-2018, 10-12-2018, 26-12-2018$ (days where the NASDAQ100 index price closes higher than its opening price) and $09-11-2018, 12-11-2018, 14-11-2018, 04-12-2018, 07-12-2018, 21-12-2018$ (days where the NASDAQ100 index price closes lower than its opening price). Further, the duration from the opening to the closing hour of the market is divided into six time-zones, as outlined in Table~\ref{tab:time_zones} for the intraday analysis. Each time zone is one hour long, except for T3 and T4, which are for one hour and fifteen minutes. This diverse process of selecting stocks and trading days ensures that the results of this study can be generalized to some extent. A comprehensive presentation of the overall findings of this study and its in-depth analysis is given in the subsequent Sec.~\ref{sec:R}.

\begin{table}[h]
\centering
\caption{Time-zones for the intraday analysis.}
\label{tab:time_zones}
\newcolumntype{C}[1]{>{\centering\arraybackslash}p{#1}}
\begin{tabular}{|C{2.2cm}|C{4.6cm}|} 
\hline
\textbf{Time-Zone} & \textbf{Timing (HH:MM:SS)} \\
\hline
T1 & 09:30:00.000 - 10:29:59.999 \\
\hline 
T2 & 10:30:00.000 - 11:29:59.999 \\
\hline
T3 & 11:30:00.000 - 12:44:59.999 \\
\hline
T4 & 12:45:00.000 - 13:59:59.999 \\
\hline
T5 & 14:00:00.000 - 14:59:59.999 \\
\hline
T6 & 15:00:00.000 - 16:00:00.000 \\
\hline
\end{tabular}
\end{table}

\section{Results}
\label{sec:R}
In this section, we analyze the intraday dynamics of high-frequency stock market order transitions by utilizing the first-order time-homogeneous discrete-time Markov chain (DTMC) model. The order transition dynamics are compared between the time-zones across HMC, MMC and LMC stocks.

In Subsec.~\ref{result:G-test}, we carry out G-test of independence to show that the sequence of orders has memory, meaning the occurrence of one event in the sequence is dependent on the occurrence of previous events. Subsec.~\ref{result:TPM_Order} describes the transition probabilities between stock market orders at different time-zones of a trading day for HMC, MMC and LMC stocks. Subsec.~\ref{result:SD_Order} shows the stationary distribution of each order at different time-zones and compares the time-zones based on the stationary distribution values using Jensen-Shannon divergence. Finally, in Subsec.~\ref{result:PCA_DBSCAN}, clustering of time-zones is carried out by using Density-Based Spatial Clustering of Applications with Noise after the dimensionality reduction of the transition probability matrices using Principal Component Analysis.

\subsection{G-test of independence}
\label{result:G-test}
The G-test statistic and the corresponding p-values of the high-frequency stock market order sequence data at different time-zones of each stocks for HMC, MMC and LMC are calculated. Table~\ref{tab:G-test} in the Appendix shows the average G-test statistic for all the trading days considered in this study. 

The high value of average G-statistic suggests a significant discrepancy between the observed data and the expected frequencies calculated under the hypothesis of independence. A very low p-value ($\ll 0.05$) indicates that the likelihood of observing such large statistic under the null hypothesis is effectively zero, leading to the rejection of the null hypothesis. This suggests that the transitions between states are not occurring randomly; rather, they exhibit dependency where the occurrence of one event in the chain is influenced by previous events. We further carried out autocorrelation analysis of the order sequence data and found that the autocorrelation values at lag 1 and lag 2 are similar and then declines at the subsequent lags. The lag 1 and lag 2 autocorrelation values are small but substantially exceeds the 1/$\sqrt N$ threshold, where $N$ represents the number of events~\cite{chatfield1978holt}. As a result, these values are statistically significant and point to the presence of short-term memory~\cite{beran1992statistical}. In this study, we focus only on the dependence between consecutive events, thus enabling the application of a first-order Markov property to high-frequency tick-by-tick order sequence data. The second-order Markov chain consideration may be carried out in future works. Under this first-order assumption, we estimate the transition probabilities using Maximum Likelihood Estimation, as presented in the next subsection.

\subsection{Transition Probability Matrix of Orders}
\label{result:TPM_Order}

The transition probabilities between the orders, AB, AA, DB, DA, FB, FA, EB, EA, CB and CA, are estimated for the 6 time-zones of each stock - 15 stocks in total, with 5 each categorized under HMC, MMC, and LMC, using Maximum Likelihood Estimation. In total, there are 1080 transition probability matrices (TPMs) of order $10\times 10$ for the 12 days considered in this study. After averaging, we have a total of 18 matrices, with 6 matrices each corresponding to HMC, MMC, and LMC. We compare these matrices between HMC, MMC, and LMC at different time-zones. Our analysis is restricted to only AB, AA, DB, DA, FB and FA as the count of orders for EB, EA, CB and CA are very low, as demonstrated in Fig.~\ref{fig:order_count} in the Appendix.

Figs.~\ref{fig:TPM_Order1} and~\ref{fig:TPM_Order2} in the Appendix show the heatmap comparison of TPMs among HMC, MMC, and LMC, respectively at different time-zones. The columns of the matrix represent the current order in the sequence whereas the rows represent the next order. The values inside the matrices represents the transition probabilities between the orders. Using these heatmaps, we compare the probabilities of dominant and physically significant order transitions at different time-zones between HMC, MMC, and LMC, as presented below in detail. 

\begin{figure}[h]
    \centering
    \begin{tabular}{ccc}
        \begin{subfigure}{0.353\textwidth}
            \includegraphics[width=\textwidth]{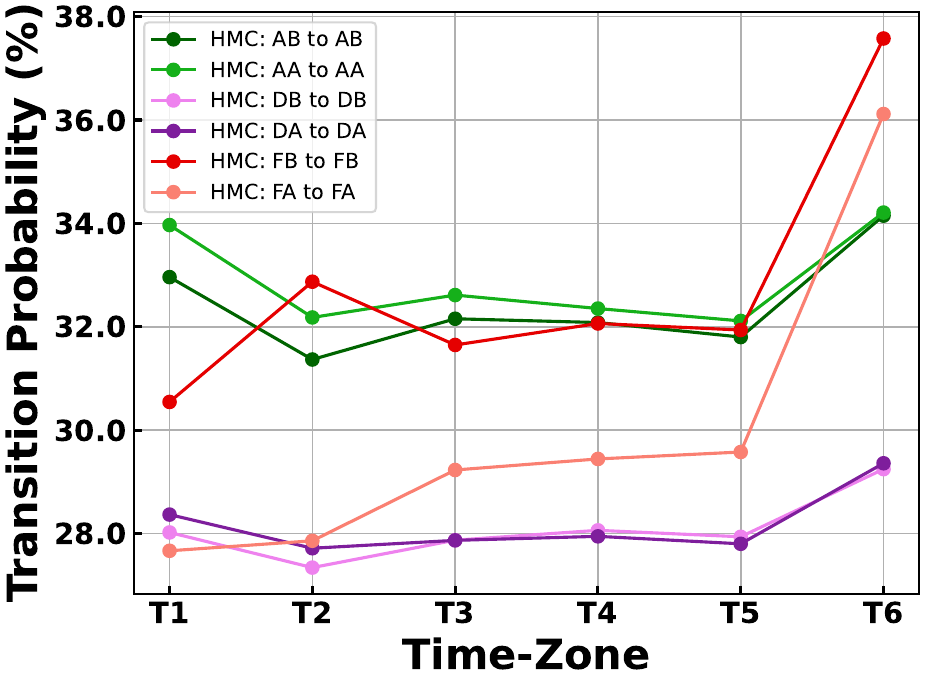}
        \end{subfigure} &
        \begin{subfigure}{0.33\textwidth}
            \includegraphics[width=\textwidth]{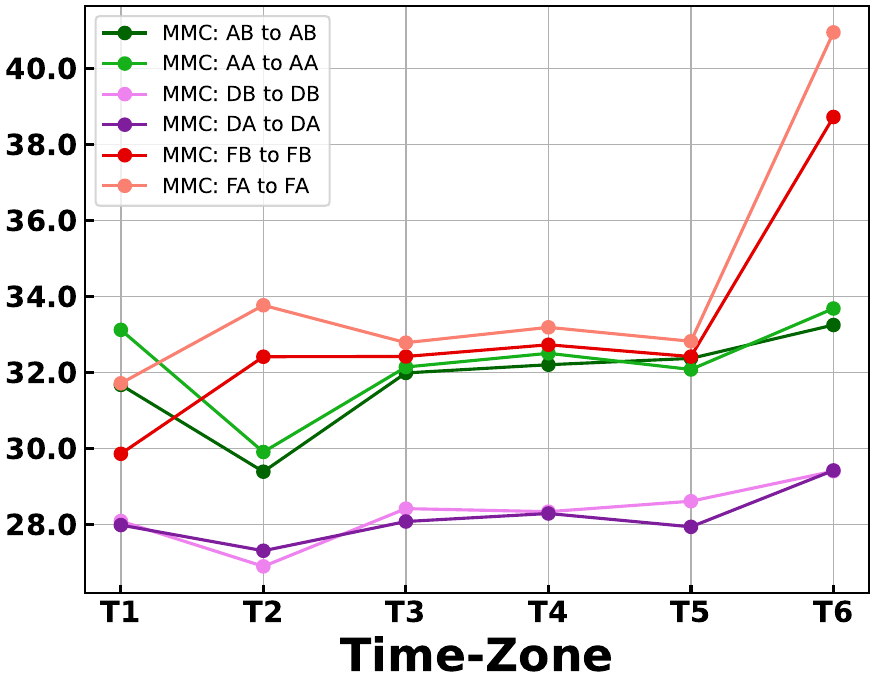}
        \end{subfigure} &
        \begin{subfigure}{0.33\textwidth}
            \includegraphics[width=\textwidth]{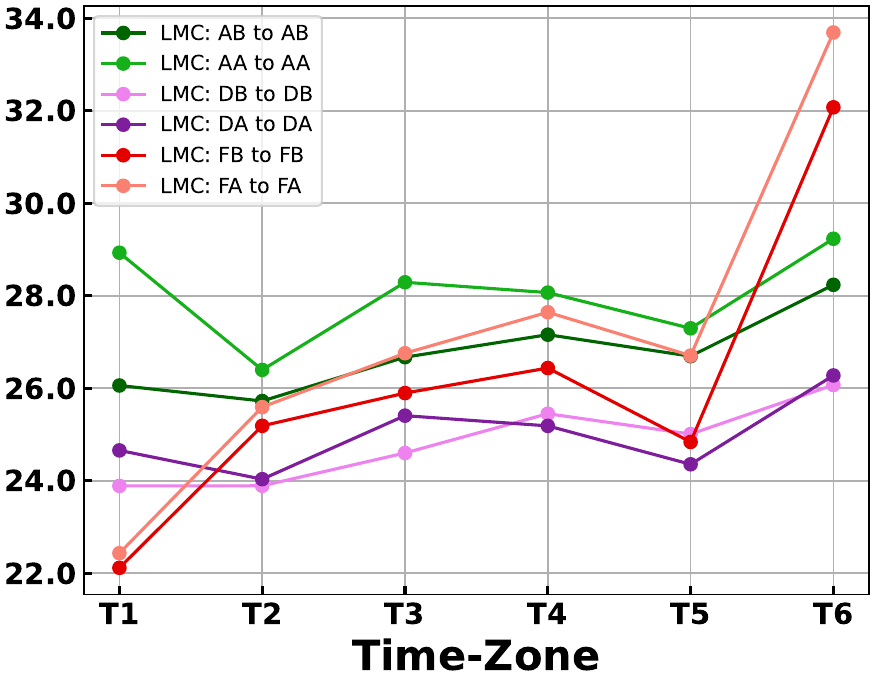}
        \end{subfigure} \\
    \end{tabular}
    \caption{Intraday variation of the DoI of orders, i.e., the transition probability from one type of order to the same type, for HMC [left], MMC [middle] and LMC [right] stocks. Different colors in the line plots represent different transitions [Green: \( \text{AB} \rightarrow \text{AB} \); Light green: \( \text{AA} \rightarrow \text{AA} \); Red: \( \text{FB} \rightarrow \text{FB} \); Orange: \( \text{FA} \rightarrow \text{FA} \); Light purple: \( \text{DB} \rightarrow \text{DB} \); Purple: \( \text{DA} \rightarrow \text{DA} \)].}
    \label{fig:dois}
\end{figure}

We use -- degree of inertia (DoI) of an order -- as defined in prior works~\cite{rabindrajit2024high,paxinou2021analyzing}, to quantify the probability that an order remains unchanged during the next transaction. DoI is derived from the diagonal entries of the TPM heatmaps, which represent the probability of an order retaining its current state. Fig.~\ref{fig:dois} shows the variation of the DoI orders, AB, AA, DB, DA, FB and FA at different time-zones for HMC, MMC, and LMC stocks.

\begin{enumerate}
    \item Intraday order transition dynamics [Refer to Fig.~\ref{fig:dois}]:\\
    During T1, the DoI of addition of buying limit order (AB) and selling limit order (AA) is relatively high as compared to other orders. In particular, the DoI of AA exceeds that of AB, suggesting a greater likelihood of consecutive selling limit order submissions. As timezone shifts from T1 to T2, we observe a decline in the DoI of AA and AB, whereas an increase in the DoI of FA and FB. This increase may be due to the increase in market liquidity due to limit order accumulations during T1, hence, market orders becoming more appealing to traders during T2. After T3, there is negligible variation in the DoI of all orders upto T5, except for LMC stocks. From T5 to T6, the DoI of all the order types increases with a significant spike in buy and sell order executions, i.e., FB and FA. This may signify that the market participants are more concern about executing the open positions. As a result, those who need to exit positions on the same day favor market orders or more aggressive limit orders (closer to the best bid/ask) to ensure execution.
    
    Order transition fluctuations during the market opening hour may make it less ideal for long-term investors to add positions. However, intraday traders may capitalize on these fluctuations, provided they analyze pre-market orders and overnight news updates. They can also take advantage from the increased liquidity demand as other traders rush to execute market orders during the closing hour.
        
    \item Differences between HMC, MMC and LMC in order transition dynamics [Refer to Fig.~\ref{fig:dois}]:\\
    The overall DoI of the orders is larger for HMC and MMC as compared to LMC stocks, which indicates that consecutive submissions of identical orders are more prevalent in HMC and MMC stocks. Consecutive submissions may imply that large orders are being split into smaller ones and placed successively to minimize market impact. Further, the DoI of bid order execution (FB) is more likely than ask order execution (FA) for HMC stocks. However, in the case of MMC and LMC stocks, the DoI of FA is more likely than that of FB. 
    
    Consecutive buys/sells in small quantities might indicate gradual order accumulation or the process of exiting a position, respectively. Once this pattern is identified, especially in higher market cap stocks, traders should trade in the same direction before the full impact becomes visible.
        
    \end{enumerate}

\begin{figure}[h]
    \centering
    \begin{tabular}{ccc}
        \begin{subfigure}{0.345\textwidth}
            \includegraphics[width=\textwidth]{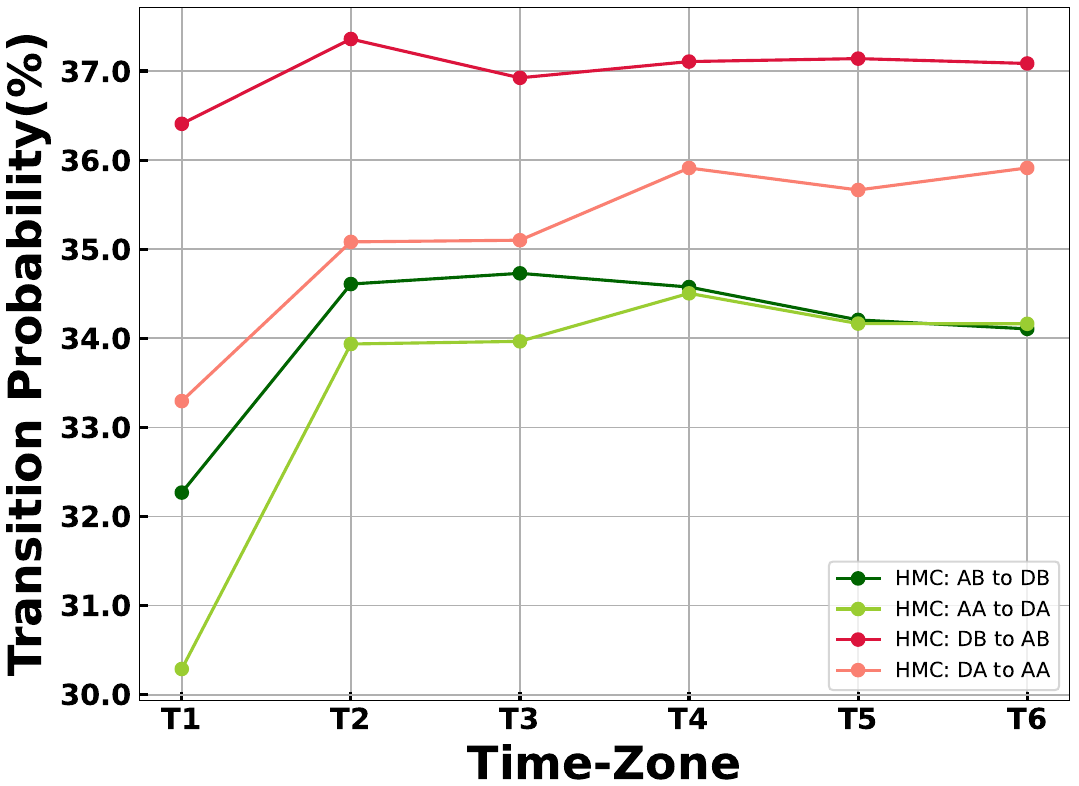}
        \end{subfigure} &
        \begin{subfigure}{0.33\textwidth}
            \includegraphics[width=\textwidth]{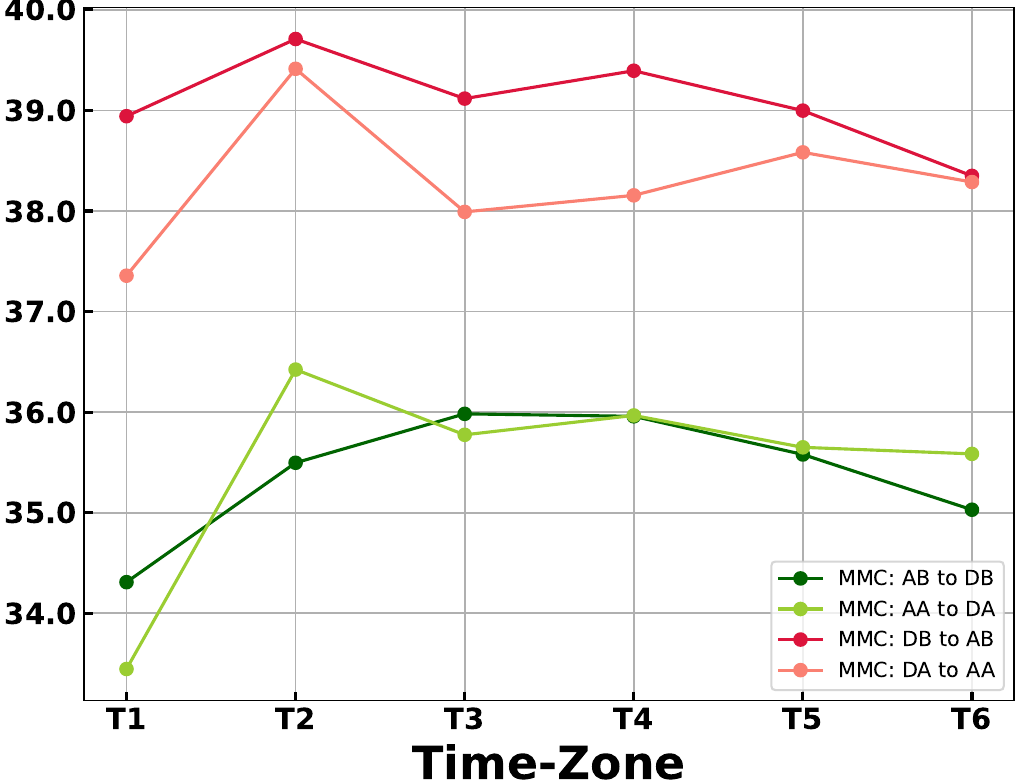}
        \end{subfigure} &
        \begin{subfigure}{0.328\textwidth}
            \includegraphics[width=\textwidth]{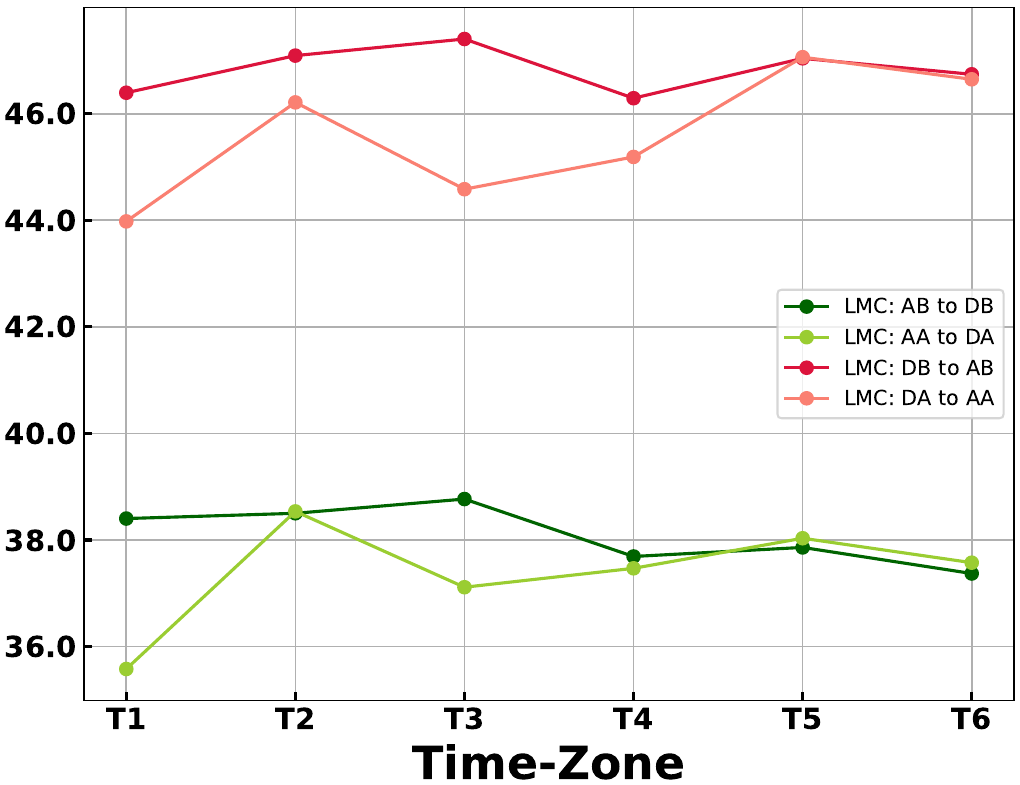}
        \end{subfigure} \\
    \end{tabular}
    \caption{Intraday variation of transition probability from addition of bid and ask limit orders to its deletion and vice-versa for HMC [left], MMC [middle] and LMC [right] stocks. Different colors in the line plots represent different transitions [Red: \( \text{DB} \rightarrow \text{AB} \); Orange: \( \text{DA} \rightarrow \text{AA} \); Green: \( \text{AB} \rightarrow \text{DB} \); Light green: \( \text{AA} \rightarrow \text{DA} \)].}
    \label{fig:AD}
\end{figure}

Fig.~\ref{fig:AD} shows the variation of the transition probability from limit order addition to its deletion and vice-versa at different time-zones for HMC, MMC and LMC stocks.

\begin{enumerate}
    \item Intraday order transition dynamics [Refer to Fig.~\ref{fig:AD}]:\\

    The transition probability from addition of limit order to its deletion, i.e., AB/AA $\rightarrow$ DB/DA, is smaller than the transition probability from deletion to addition of limit order, i.e., DB/DA $\rightarrow$ AB/AA. Further, the transition probability from AB/AA $\rightarrow$ DB/DA and vice-versa increases as the timezone shifts from T1 to T2. This indicates that the probability of limit order adjustments through additions and deletions increases from T1 to T2. The increase limit order modifications after the opening hour could mean that initial orders placed in anticipation of certain moves during T1 failed to occur, leading to market correction during T2.
    
    Therefore, traders should remain proactive to adjust limit orders to match the new trends during the market correction period. 

    \item Differences between HMC, MMC and LMC in order transition dynamics [Refer to Fig.~\ref{fig:AD}]:\\
    The overall transition probability from  AB/AA $\rightarrow$ DB/DA and vice-versa is highest for LMC, followed by MMC and HMC, indicating higher likelihood of limit order modifications in LMC stocks, similar to the finding in Ref.~\cite{foucault1999order}. Further, the difference between the transition probability from DB/DA $\rightarrow$ AB/AA and AB/AA $\rightarrow$ DB/DA is highest for LMC stocks. This indicates that the probability of immediate addition after deletion of a limit order is more prominent as compared to the probability of immediate deletion after addition for lower market cap stocks. As market cap increases, this prominence decreases.
    
    Given the frequent limit order modifications in lower market cap stocks, traders should focus on trading during higher liquidity periods to minimize the impact of these modifications and reduce slippage. They may also reduce position sizes to limit exposure and use stop-limit orders instead of market orders to ensure exits within acceptable price ranges.
    
    \end{enumerate}

\begin{figure}[h]
    \centering
    \begin{tabular}{ccc}
        \begin{subfigure}{0.35\textwidth}
            \includegraphics[width=\textwidth]{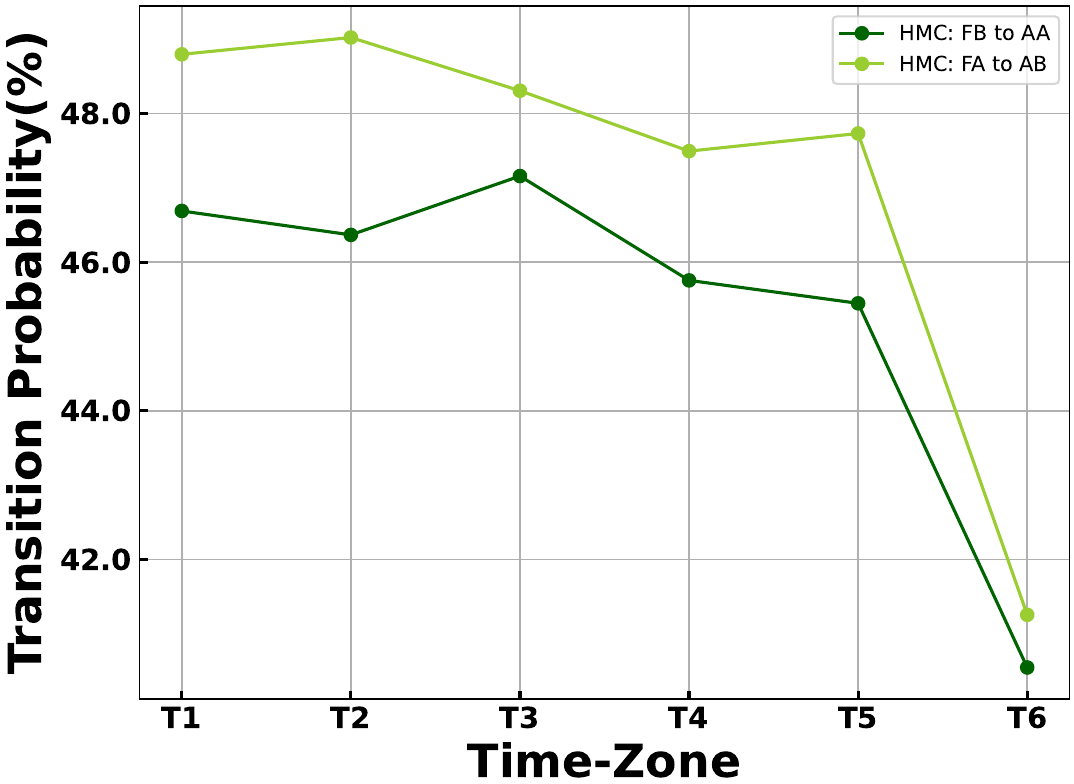}
        \end{subfigure} &
        \begin{subfigure}{0.33\textwidth}
            \includegraphics[width=\textwidth]{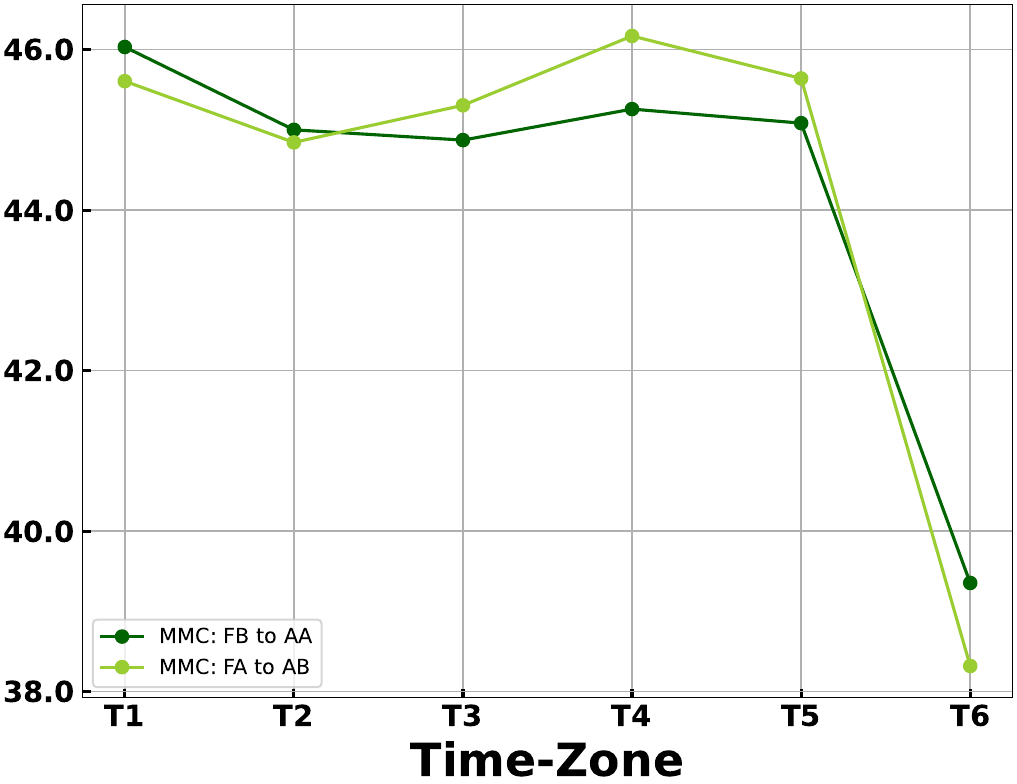}
        \end{subfigure} &
        \begin{subfigure}{0.33\textwidth}
            \includegraphics[width=\textwidth]{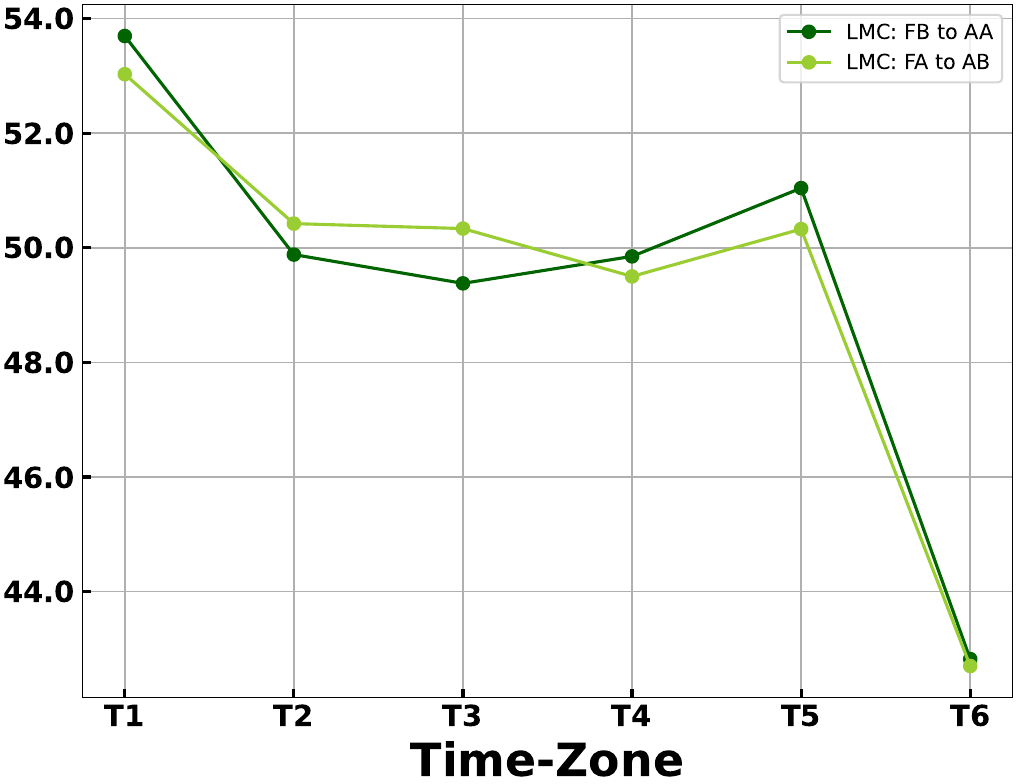}
        \end{subfigure} \\
    \end{tabular}
    \caption{Intraday variation of transition probability from execution (in full quantity) to the addition of bid and ask limit orders for HMC [left], MMC [middle] and LMC [right] stocks. Different colors in the line plots represent different transitions [Green: \( \text{FB} \rightarrow \text{AA} \); Light green: \( \text{FA} \rightarrow \text{AB} \)].}
    \label{fig:FA}
\end{figure}

Fig.~\ref{fig:FA} shows the variation of the transition probability from order execution to limit order addition at different time-zones for HMC, MMC and LMC stocks. 

\begin{enumerate}
    \item Intraday order transition dynamics [Refer to Fig.~\ref{fig:FA}]:\\
   
    The transition probabilities from order execution to limit order addition, i.e., \( \text{FB} \rightarrow \text{AA} \) and \( \text{FA} \rightarrow \text{AB} \), significantly decrease as the timezone shifts from T5 to T6. This indicates a sharp decline in the addition of buy and sell limit orders following sell and buy order executions, respectively. As traders prioritize closing positions over initiating new ones, limit orders are generally not preferred. At the market closing hour, traders often favor immediate execution certainty, which limit orders cannot guarantee due to their price-specific nature. Furthermore, there is insufficient time to revise the unexecuted limits. These factors -- execution risk and time constraints -- reduce the appeal of limit orders in the final trading hour.
    
    Therefore, as the addition of new limit orders drops sharply during the closing hour, traders should avoid placing passive limit orders (limit order far away from the current best price), as the likelihood of these orders remaining unexecuted is high.
    
    \item Differences between HMC, MMC and LMC in order transition dynamics [Refer to Fig.~\ref{fig:FA}]:\\
    
    The transition probabilities, \( \text{FB} \rightarrow \text{AA} \) and \( \text{FA} \rightarrow \text{AB} \) are highest for LMC stocks. This means that the probability of submitting a sell limit order after a buy order execution or a buy limit order after a sell order execution is comparatively larger for LMC stocks. This observation may be due to LMC stocks' lower liquidity with wide bid-ask spreads~\cite{foucault1999order}, where a buy or sell order execution moves the stock price sharply. As the price movement have mean reversion tendencies, where prices revert to its recent average after sharp moves, a trader may expect that the stock price will rise after the buy order execution, so they place a sell limit order to capture quick gains (e.g., "I bought at \$90, now I'll sell if the price rises to \$100"). Conversely, if they sell first, they might expect a fall in the stock price and place a buy limit order to re-enter at a lower discounted price (e.g., "I sold at \$100, now I'll buy if the price falls to \$90"). Further, we also observed that the transition probability, \( \text{FA} \rightarrow \text{AB} \) is higher than \( \text{FB} \rightarrow \text{AA} \) for HMC stocks. This means that it is more likely to submit buy limit order after sell order execution, indicating that traders are more comfortable to re-enter the market after exiting in the case of HMC stocks.

    \end{enumerate}

After analyzing and comparing the intraday variation of dominant and physically significant order transitions between  HMC, MMC and LMC, we estimate the stationary distributions of each state and compare between different time-zones using Jensen-Shannon Divergence, as presented below in detail. 


\subsubsection{Stationary Distribution of orders}
\label{result:SD_Order}
The stationary distribution of the DTMC refers to the long-run probability distribution that remains unchanged as time progresses. The Markov chain for each time zone of a stock is found to be ergodic and irreducible. As a result, the stationary distribution can be calculated for each of them. Table~\ref{tab:SD_Order} in the Appendix shows the stationary distribution of each order type for HMC, MMC and LMC stocks at different time-zones. In order to compare the stationary distribution values, we calculate the Jensen-Shannon divergence (JSD) between the distributions of each time-zones for HMC, MMC and LMC. The JSD values that are closest to zero represent similar distributions, whereas the highest JSD values indicate the most significant differences in distributions. Tables~\ref{tab:JSD_order_large},~\ref{tab:JSD_order_mid} and~\ref{tab:JSD_order_small} show the JSD values comparing the time-zones for HMC, MMC and LMC stocks, respectively. 

\begin{table}[H]
   \centering
   \caption{Jensen-Shannon divergence between stationary distributions of different time-zones for HMC stocks.}
   \label{tab:JSD_order_large}
   \begin{tabular}{|c|c|c|c|c|c|c|}
   \cline{1-2}
   T1 & 0.0     \\
   \cline{1-3}
   T2 & 0.0182 & 0.0    \\
   \cline{1-4}
   T3 & 0.0203 & 0.0056 & 0.0   \\
   \cline{1-5}
   T4 & 0.0225 & 0.0069 & 0.0045 & 0.0    \\
   \cline{1-6}
   T5 & 0.0190 & 0.0050 & 0.0048 & 0.0050 & 0.0    \\
   \cline{1-7}
   T6 & 0.0157 & 0.0218 & 0.0252 & 0.0265 & 0.0220 & 0.0    \\
   \hline
       & T1     & T2     & T3     & T4     & T5     & T6     \\
       \hline
   \end{tabular}
\end{table}

Table~\ref{tab:JSD_order_large} shows that the pairs T3 and T4 (0.0045), T3 and T5 (0.0048), and T4 and T5 (0.0050) for HMC stocks have low JSD, indicating similarity in their distributions. Whereas, the pairs T4 and T6 (0.0265), T3 and T6 (0.0252), and T5 and T6 (0.0220) have relatively higher divergence, highlighting their dissimilarity. Particularly, T6 consistently appears in these most dissimilar pairings, indicating that the distribution of T6 is distinctly different from T3, T4, and T5. Similarly, Table~\ref{tab:JSD_order_small} shows minimal divergence for the time-zones, T2, T3, T4, and T5 for LMC stocks. T6 is also distinctly different from other time-zones. This trend continues for MMC stocks as shown in Table~\ref{tab:JSD_order_mid}, where T2, T3, T4, and T5 also demonstrate low divergence. However, for MMC stocks, the pairs T1 and T3 (0.0729), T1 and T4 (0.0744), and T1 and T5 (0.0728) have higher divergence, indicating that the distribution of T1 is distinctly different from T3, T4, and T5, unlike the distinction of T6 for HMC and LMC stocks.

\begin{table}[H]
   \centering
   \caption{Jensen-Shannon divergence between stationary distributions of different time-zones for MMC stocks.}
   \label{tab:JSD_order_mid}
   \begin{tabular}{|c|c|c|c|c|c|c|}
   \cline{1-2}
   T1 & 0.0     \\
   \cline{1-3}
   T2 & 0.0145 & 0.0    \\
   \cline{1-4}
   T3 & 0.0729 & 0.0677 & 0.0   \\
   \cline{1-5}
   T4 & 0.0744 & 0.0689 & 0.0033 & 0.0    \\
   \cline{1-6}
   T5 & 0.0728 & 0.0675 & 0.0024 & 0.0041 & 0.0    \\
   \cline{1-7}
   T6 & 0.0715 & 0.0663 & 0.0234 & 0.0247 & 0.0226 & 0.0    \\
   \hline
       & T1     & T2     & T3     & T4     & T5     & T6     \\
   \hline
   \end{tabular}
\end{table}

\begin{table}[H]
   \centering
   \caption{Jensen-Shannon divergence between stationary distributions of different time-zones for LMC stocks.}
   \label{tab:JSD_order_small}
   \begin{tabular}{|c|c|c|c|c|c|c|}
   \cline{1-2}
   T1 & 0.0     \\
   \cline{1-3}
   T2 & 0.0142 & 0.0    \\
   \cline{1-4}
   T3 & 0.0136 & 0.0093 & 0.0   \\
   \cline{1-5}
   T4 & 0.0154 & 0.0068 & 0.0072 & 0.0    \\
   \cline{1-6}
   T5 & 0.0159 & 0.0068 & 0.0096 & 0.0057 & 0.0    \\
   \cline{1-7}
   T6 & 0.0344 & 0.0326 & 0.0362 & 0.0347 & 0.0318 & 0.0    \\
   \hline
       & T1     & T2     & T3     & T4     & T5     & T6     \\
   \hline
   \end{tabular}
\end{table}

 These findings show that the long term order placement probabilities during midday trading hours exhibit similarity across HMC, MMC, LMC stocks, suggesting a stable equilibrium in order submission behavior by traders. However, differences arise during market opening and closing hour: MMC stocks display distinct order submission dynamics during opening hours, whereas HMC and LMC stocks exhibit unique dynamics during closing sessions, as compared to the rest of the trading hours. Therefore, traders should take precautionary measures while dealing with HMC and LMC stocks during the closing hour due to potential sudden and inconsistent order transitions. Similar caution should also be applied during the opening hour when trading MMC stocks.
 
 We further perform clustering of the TPMs corresponding to different time-zones using the DBSCAN technique, following dimensionality reduction of the TPMs through Principal Component Analysis. The detailed results are provided in the following subsection.


\subsubsection{PCA and DBSCAN}
\label{result:PCA_DBSCAN}

\begin{figure}[h]
    \centering
    \begin{subfigure}{0.41\textwidth}
        \centering
        \includegraphics[width=\textwidth]{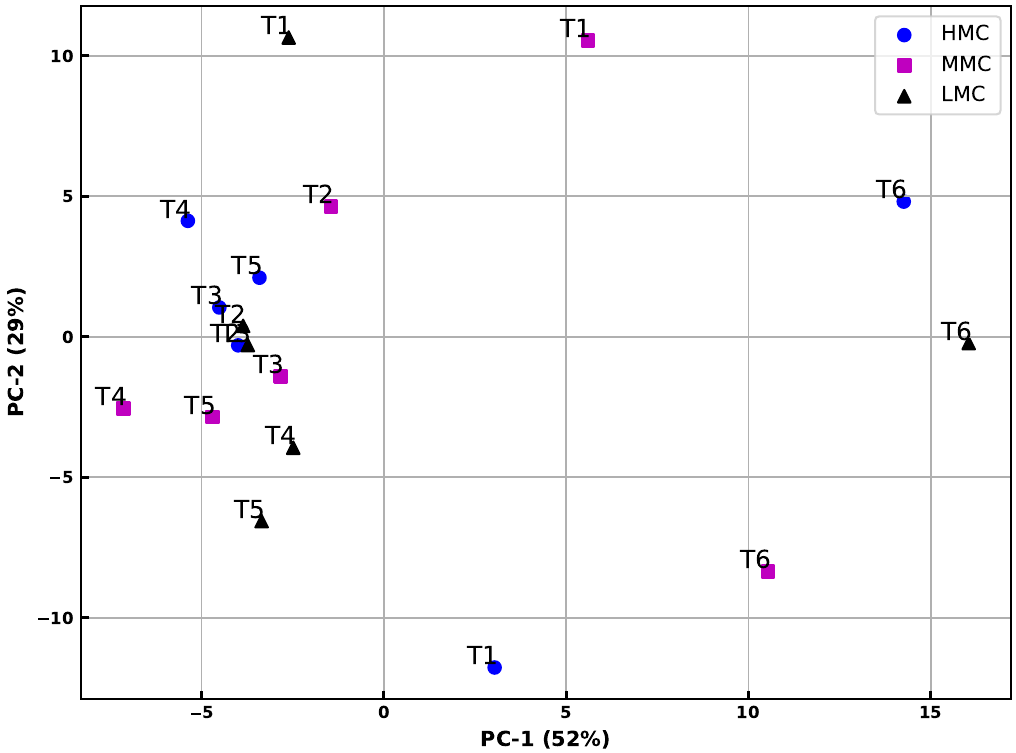} 
        \caption{2-D plot of time-zones after dimensional reduction of TPMs using PCA.}
        \label{fig:PCA}
    \end{subfigure}
    \hspace{0.2cm}
    \begin{subfigure}{0.4\textwidth} 
        \centering
        \includegraphics[width=\textwidth]{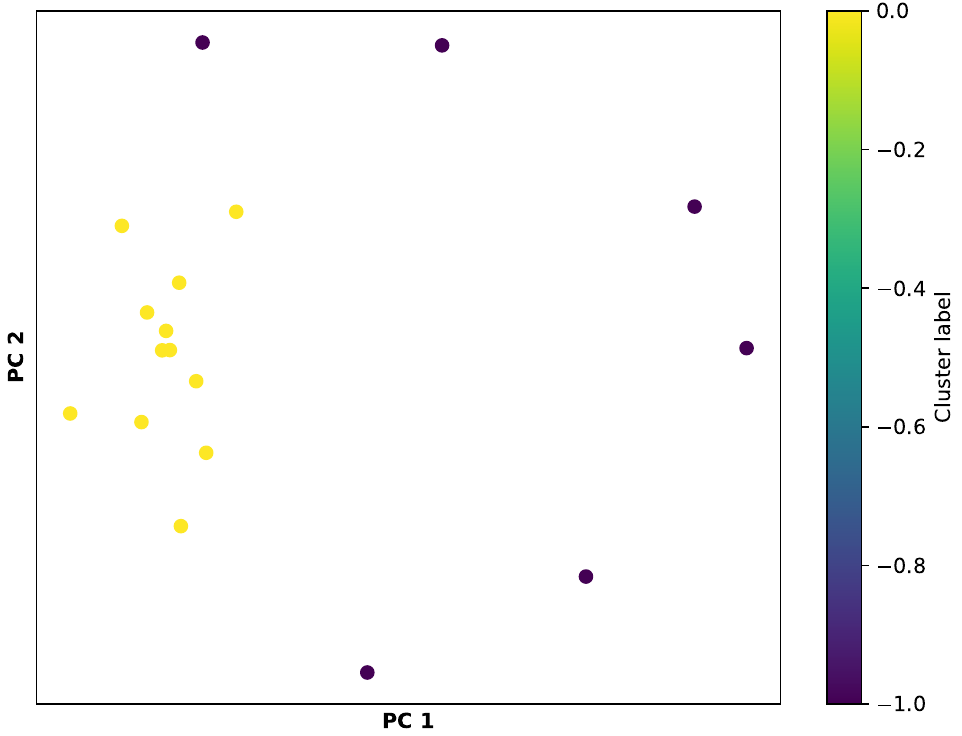}
        \caption{Clustering of different time-zones with DBSCAN.}
        \label{fig:DBSCAN}
    \end{subfigure}
    \caption{Clustering of different time-zones for HMC, MMC and LMC stocks.}
    \label{fig:clustering}
\end{figure}

In the last Subsec.~\ref{result:TPM_Order}, we have observed that the order transition dynamics at different timezones are different from each other, particularly the opening and closing hour timezones. We now carry out density-based spatial clustering of applications with noise (DBSCAN) of the TPMs of different time-zones for  HMC, MMC and LMC after reducing the dimension of the matrices with prinicipal component analysis (PCA). The $10\times 10$ TPM is first converted into into an array ($1\times 100$). With PCA, the $1\times 100$ array is reduced to $1\times 2$ by considering only the first two principal components, PC-1 and PC-2. After the dimension reduction, we obtained an $18\times 2$ array for the 18 TPMs. The first two PCs, i.e., the $18\times 2$ array, 
contains information more than $80\%$ of the original $18\times 100$ array. Hence, the two-dimensional scatter-plot as shown in Fig.~\ref{fig:PCA} is a very good approximation to the original TPMs dataset.

Clustering is carried out on the $18\times 2$ data using DBSCAN method, with the eps parameter set at $3.95$. Fig.~\ref{fig:DBSCAN} shows the clustering of different time-zones for HMC, MMC and LMC stocks. time-zones, T2, T3, T4 and T5 form a cluster. This indicates that the order transition dynamics from 10:30 to 2:00 are similar, irrespective of the market cap. However, T1 and T6 does not form a part of the cluster, indicating a different order transition dynamics in the morning and closing market hour from the dynamics of midday (T2 - T5) market hours. These findings are found to be inconsistent with this study~\cite{admati1988theory}, which found morning and closing hours as unique clustering points while attempting to explain the high NYSE
volume at open and close.


\section{Conclusions}
\label{sec:Conc}

In this study, we used a first-order time-homogeneous discrete time Markov chain model to analyze the intraday order transition dynamics across three categories -- High Market Cap (HMC), Medium
Market Cap (MMC), and Low Market Cap (LMC) -- listed on the NASDAQ100.

The study reveals that the degree of inertia (DoI) of orders, as documented in Refs.~\cite{ellul2007order, biais1995empirical,griffiths1998information} as the positive serial correlation in order flow, is higher for limit orders during the opening hour. At the subsequent trading hour, the DoI of limit order decreases while it increases for market orders. At the same time, we also found that the probability of limit order modifications through additions and deletions increases. The order transition activity remains stable during the mid-hours. As the closing hour approaches, we observed a rapid rise in the probability of consecutive order executions, consistent with prior findings~\cite{roth1988deadline,harris1998optimal,hollifield2004empirical} which suggested that traders become aggressive to execute orders at the final trading moments. This rise at the final hour was accompanied by a significant decline in the placement of buy/sell limit orders following sell/buy order executions, respectively, indicating less interest from traders in placing limit orders. 

In terms of the differences in order transitions between HMC, MMC and LMC stocks, the DoI of orders was found to be more dominant in HMC stocks with buying activity outweighing the selling activity. On the other hand, the probability of limit order modifications is greater LMC stocks. In addition, the probability to submit a sell/ buy limit order after a buy/sell order execution, as studied in Ref.~\cite{parlour1998price}, is comparatively larger in LMC stocks. The study also finds that the long term order placement probabilities during midday trading hours exhibit similarity across HMC, MMC, LMC stocks, suggesting a stable equilibrium in order submission behavior by traders. However, the closing hour exhibits distinct behavior for HMC and LMC stocks, while the opening hour differs from the other time-zones for MMC stocks. Finally, order transition activity during midday trading hours is also found to be clustered for all the stocks, a similar finding as in Ref.~\cite{ellul2007order}, except for opening and closing hours. 

These findings may be used by the market participants to refine their order placement strategies according to specific trading hours and market cap of stocks. The present study may be extended to study the transitions in intraday limit order price changes across various market cap. By broadening the analysis, we can better understand the intraday price changes so that one may capitalize on short-term price movements. 

\begin{acknowledgments}
We extend our gratitude to Chris Bartlett, Jodhie Cabarles, and the technical support staff of Algoseek~\cite{Algo} for generously providing the data and offering assistance with data preprocessing necessary for our analysis. I would also like to thank the Director of our institute for allocating doctoral research fellowship.
\end{acknowledgments}

\section*{Data Availability Statement}


The data that support the findings of this study are available from Algoseek~\cite{Algo}. Restrictions apply to the availability of these data, which were used under license for this study. Data are available from the authors upon reasonable request and with the
permission of Algoseek~\cite{Algo}.

\section*{References}
\bibliographystyle{elsarticle-num} 
\bibliography{References}

\appendix

\section{Results}

\begin{figure}[H]
    \centering
    \includegraphics[width=10.5cm]{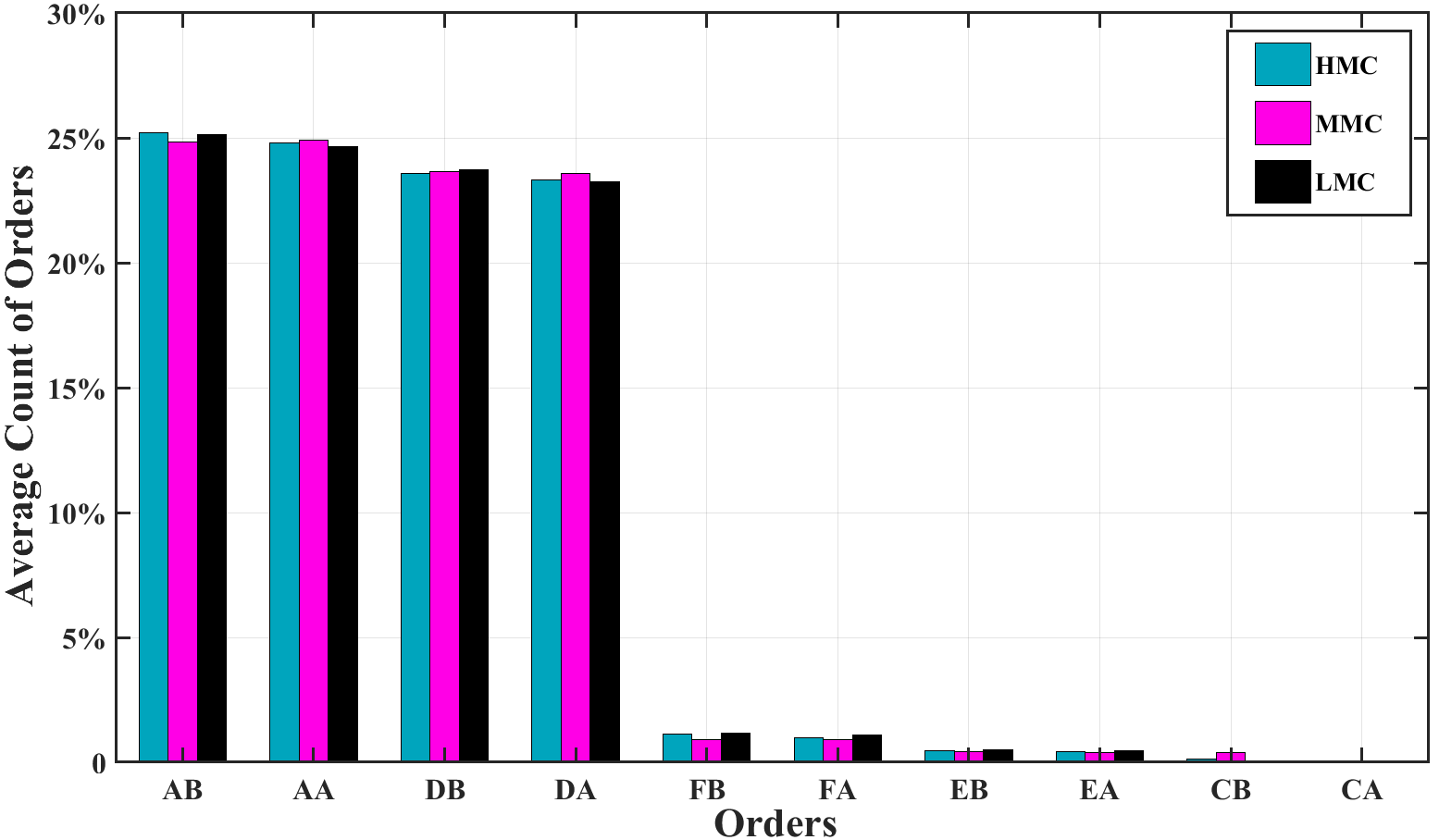}
    \caption{Average count of order submissions for HMC, MMC and LMC stocks.}
    \label{fig:order_count}
\end{figure}

\begin{figure}[H]
    \centering
    \begin{tabular}{ccc}
        \begin{subfigure}{0.3\textwidth}
            \includegraphics[width=\textwidth]{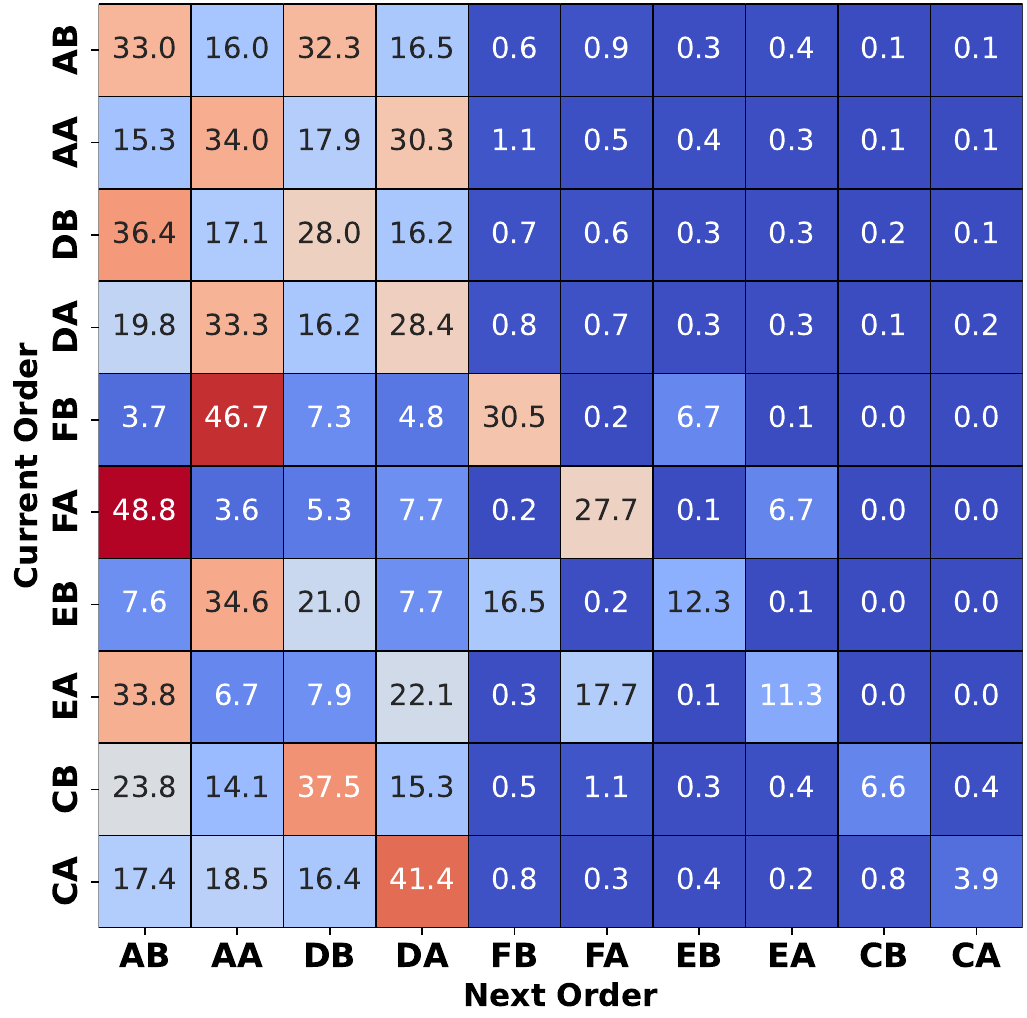}
            \caption{HMC-T1}
        \end{subfigure} &
        \begin{subfigure}{0.3\textwidth}
            \includegraphics[width=\textwidth]{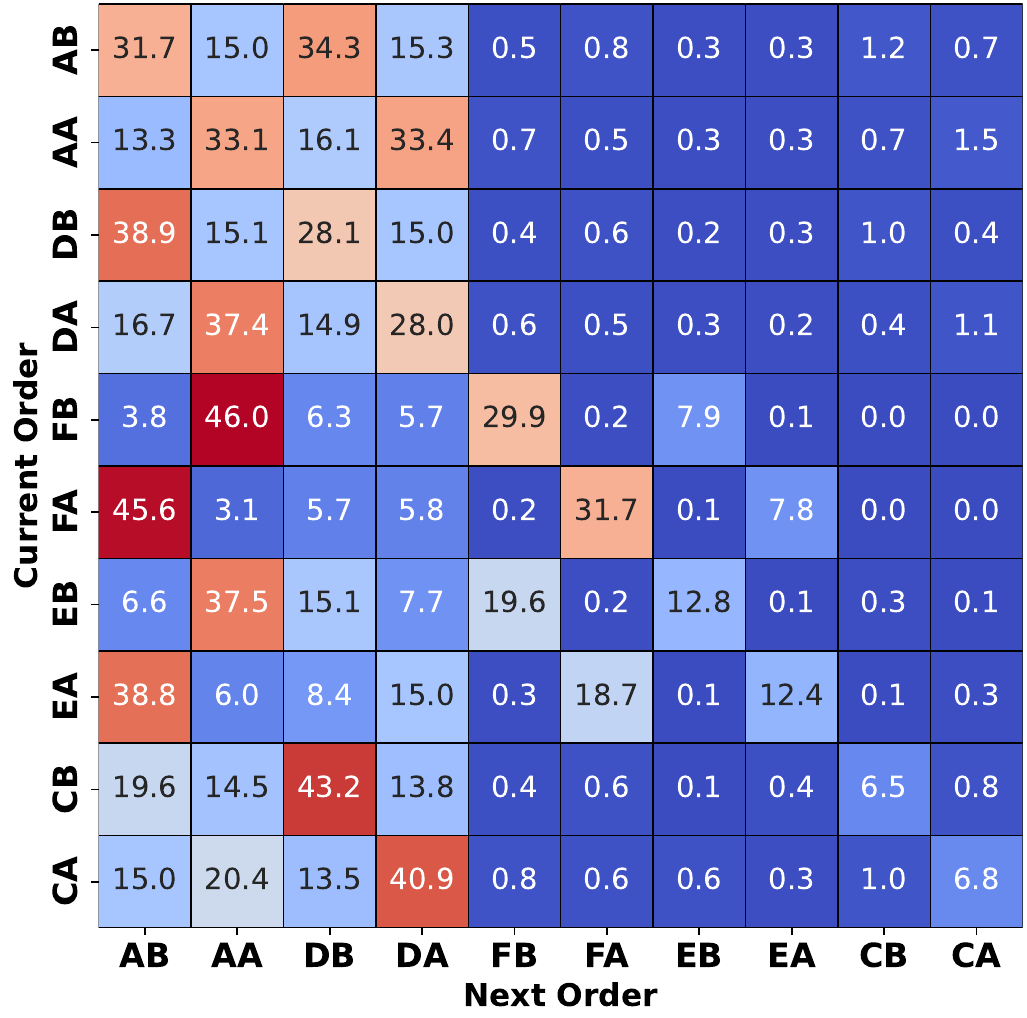}
            \caption{MMC-T1}
        \end{subfigure} &
        \begin{subfigure}{0.3\textwidth}
            \includegraphics[width=\textwidth]{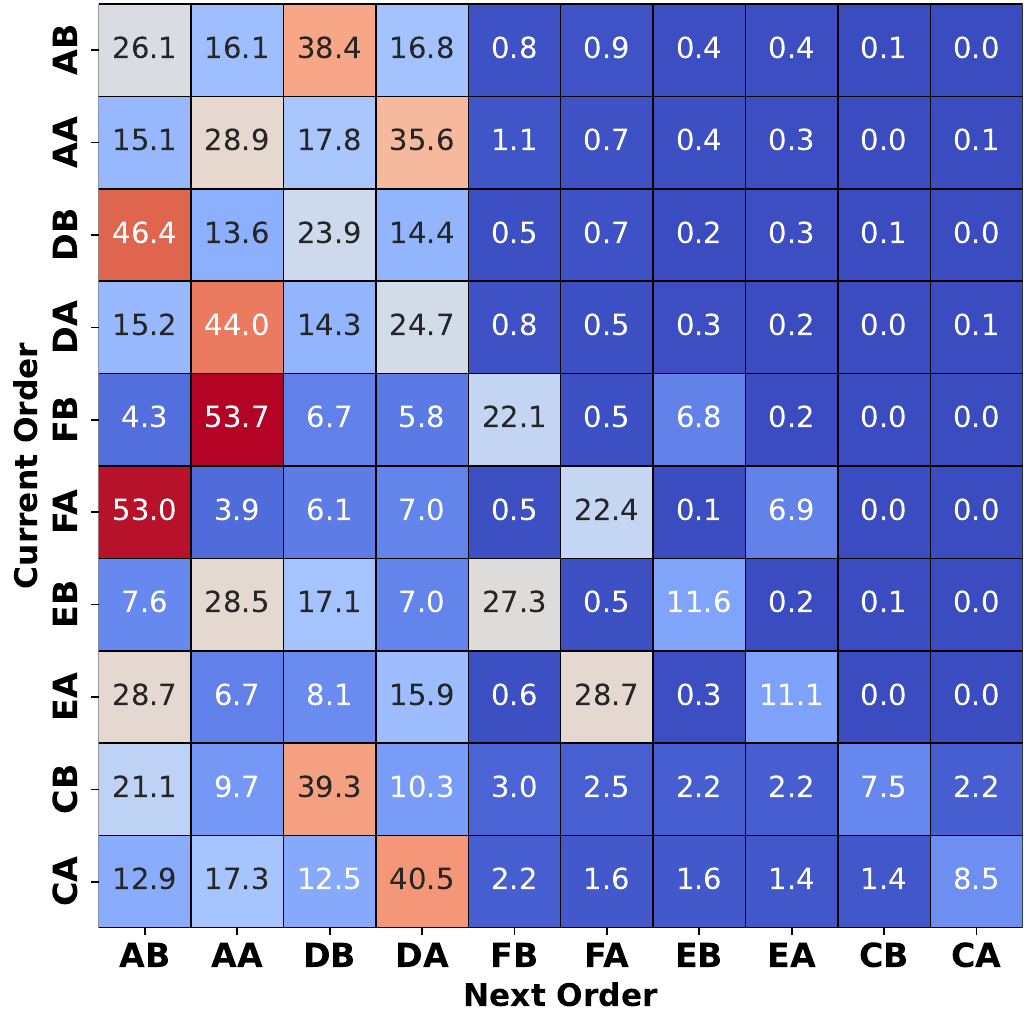}
            \caption{LMC-T1}
        \end{subfigure} \\

        \begin{subfigure}{0.3\textwidth}
            \includegraphics[width=\textwidth]{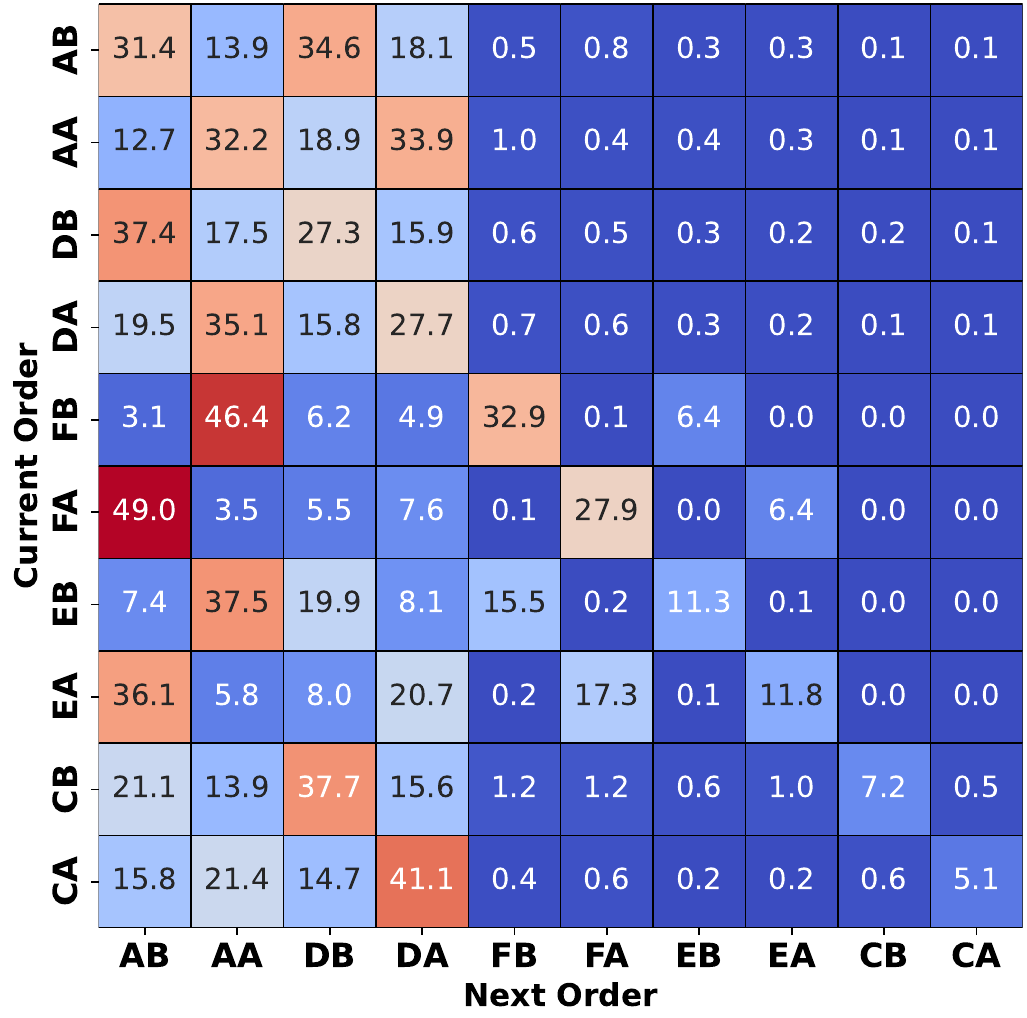}
            \caption{HMC-T2}
        \end{subfigure} &
        \begin{subfigure}{0.3\textwidth}
            \includegraphics[width=\textwidth]{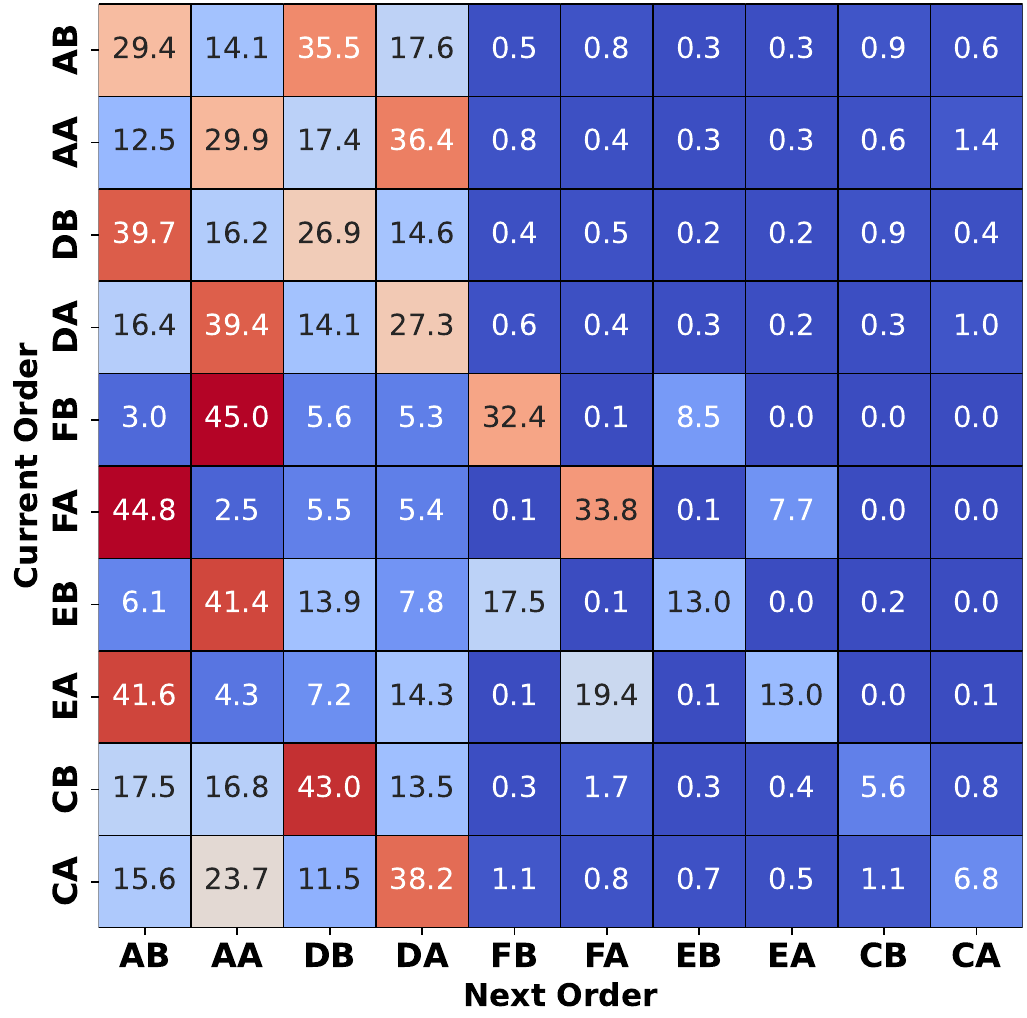}
            \caption{MMC-T2}
        \end{subfigure} &
        \begin{subfigure}{0.3\textwidth}
            \includegraphics[width=\textwidth]{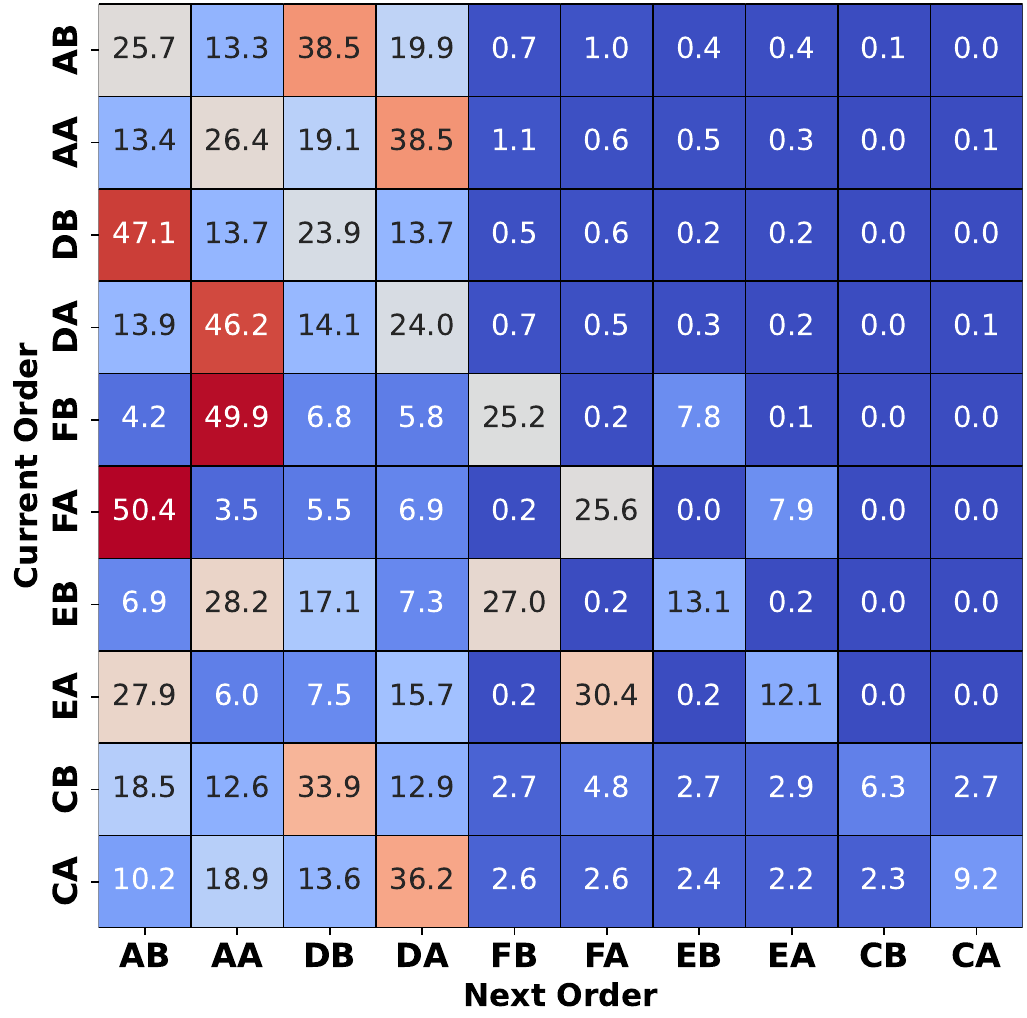}
            \caption{LMC-T2}
        \end{subfigure} \\

        \begin{subfigure}{0.3\textwidth}
            \includegraphics[width=\textwidth]{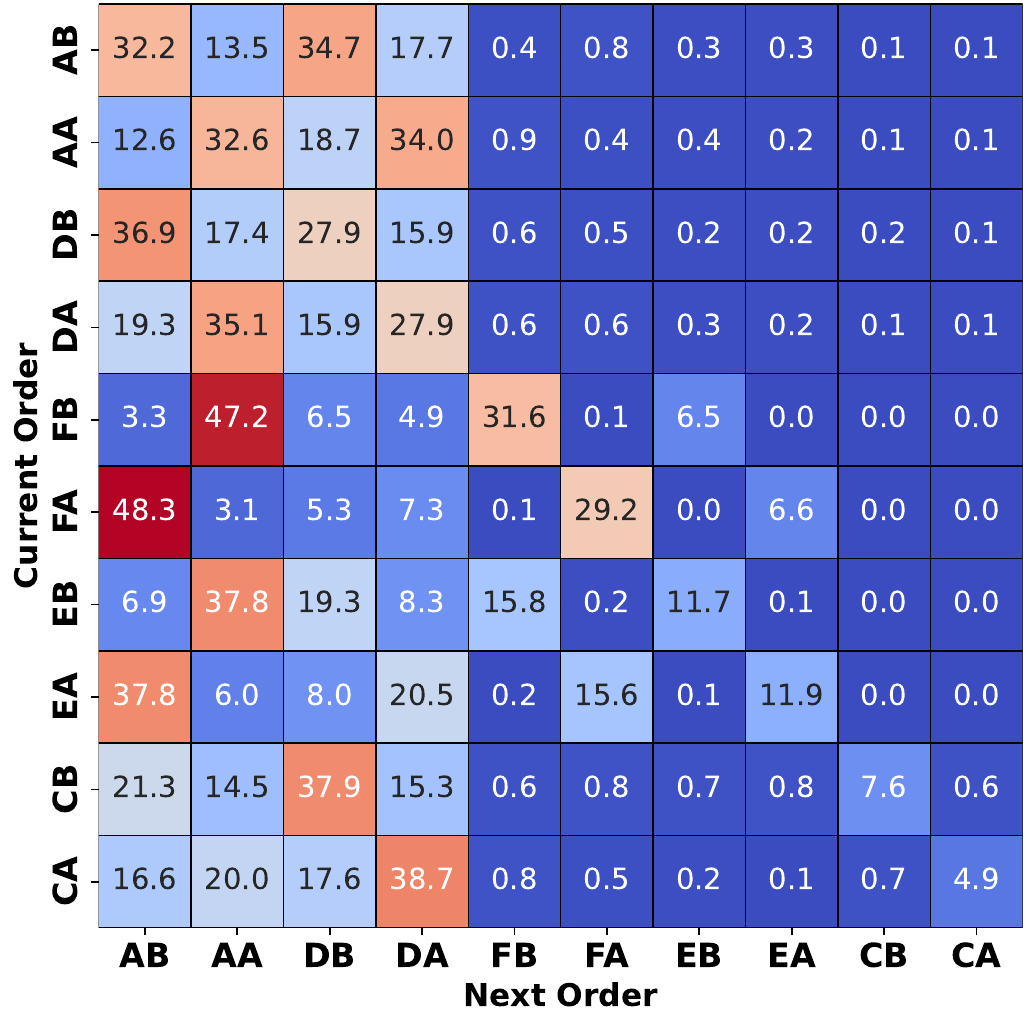}
            \caption{HMC-T3}
        \end{subfigure} &
        \begin{subfigure}{0.3\textwidth}
            \includegraphics[width=\textwidth]{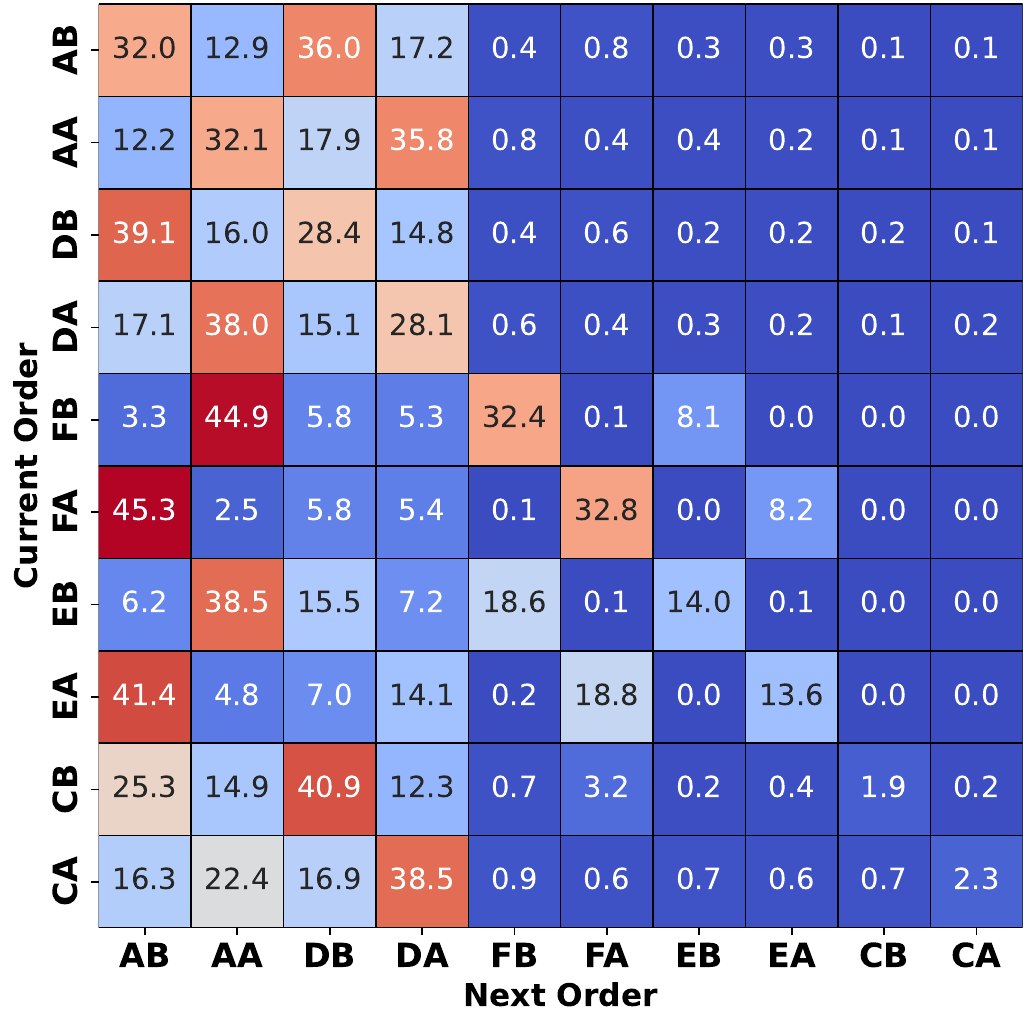}
            \caption{MMC-T3}
        \end{subfigure} &
        \begin{subfigure}{0.3\textwidth}
            \includegraphics[width=\textwidth]{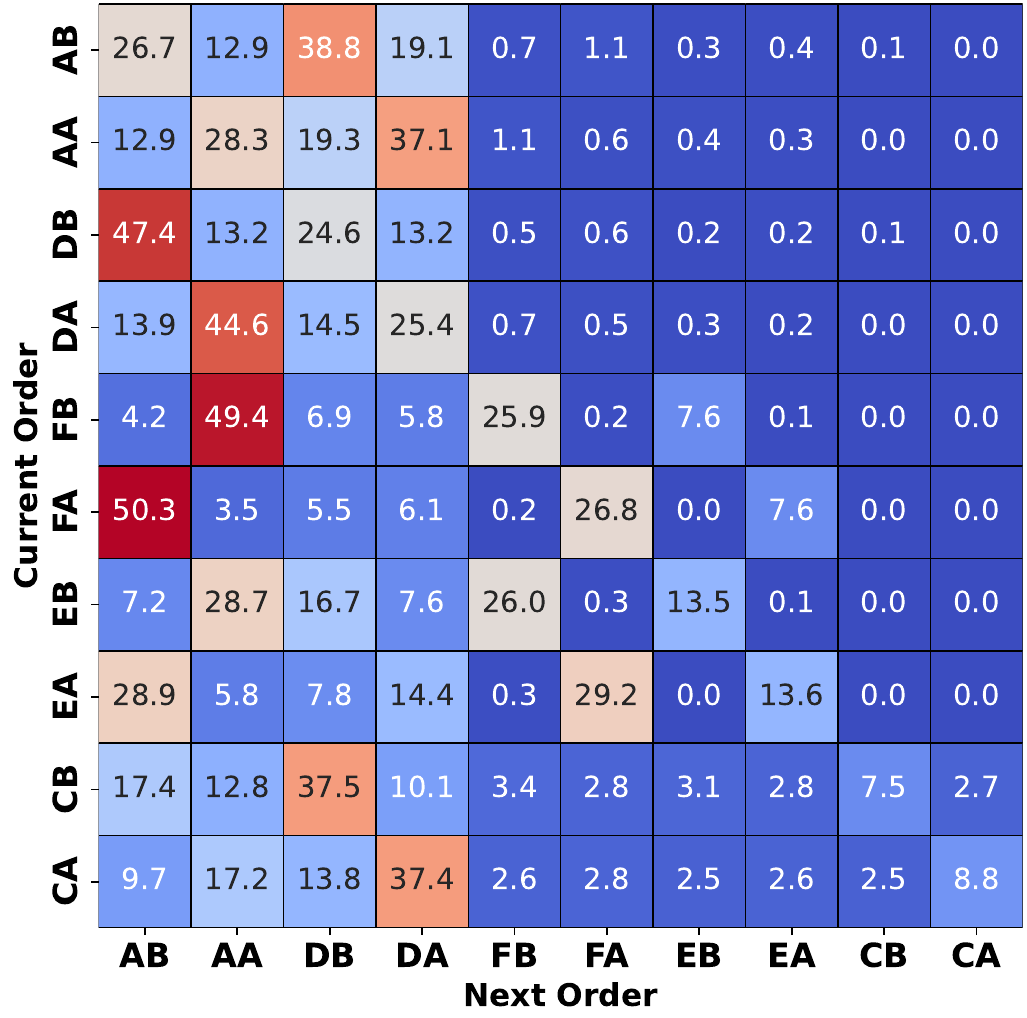}
            \caption{LMC-T3}
        \end{subfigure} \\
    \end{tabular}
    \caption{TPMs of stock market orders during different time-zones of a trading day for HMC, MMC and LMC stocks.}
    \label{fig:TPM_Order1}
\end{figure}

\begin{figure}[H]
    \centering
    \begin{tabular}{ccc}
        \begin{subfigure}{0.3\textwidth}
            \includegraphics[width=\textwidth]{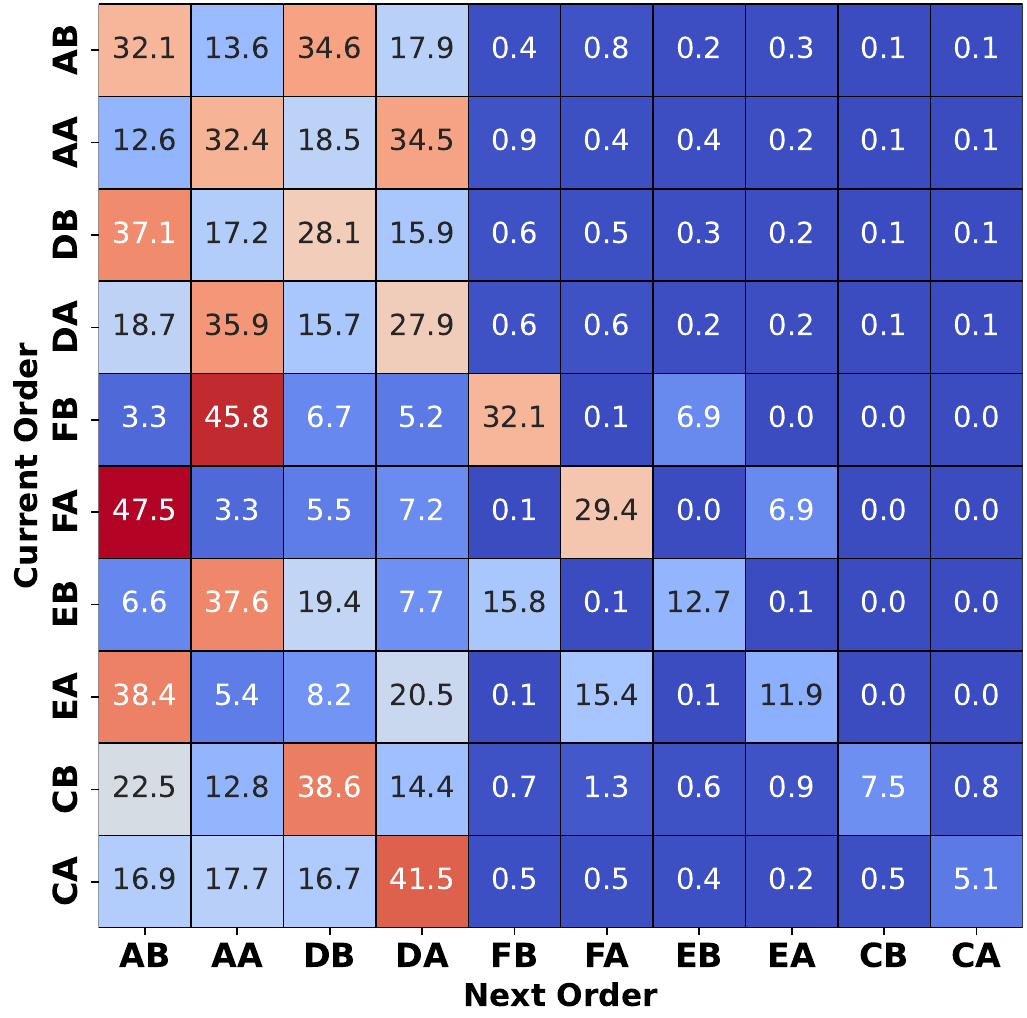}
            \caption{HMC-T4}
        \end{subfigure} &
        \begin{subfigure}{0.3\textwidth}
            \includegraphics[width=\textwidth]{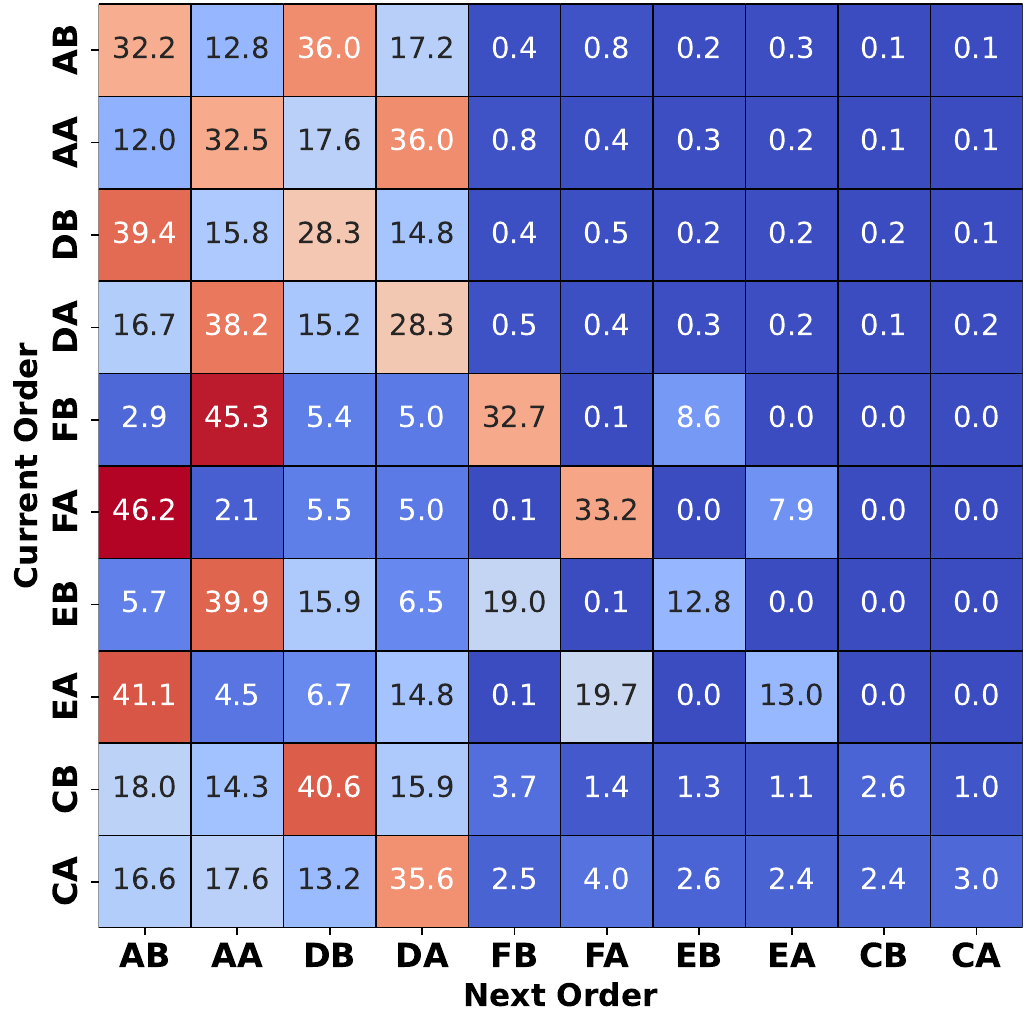}
            \caption{MMC-T4}
        \end{subfigure} &
        \begin{subfigure}{0.3\textwidth}
            \includegraphics[width=\textwidth]{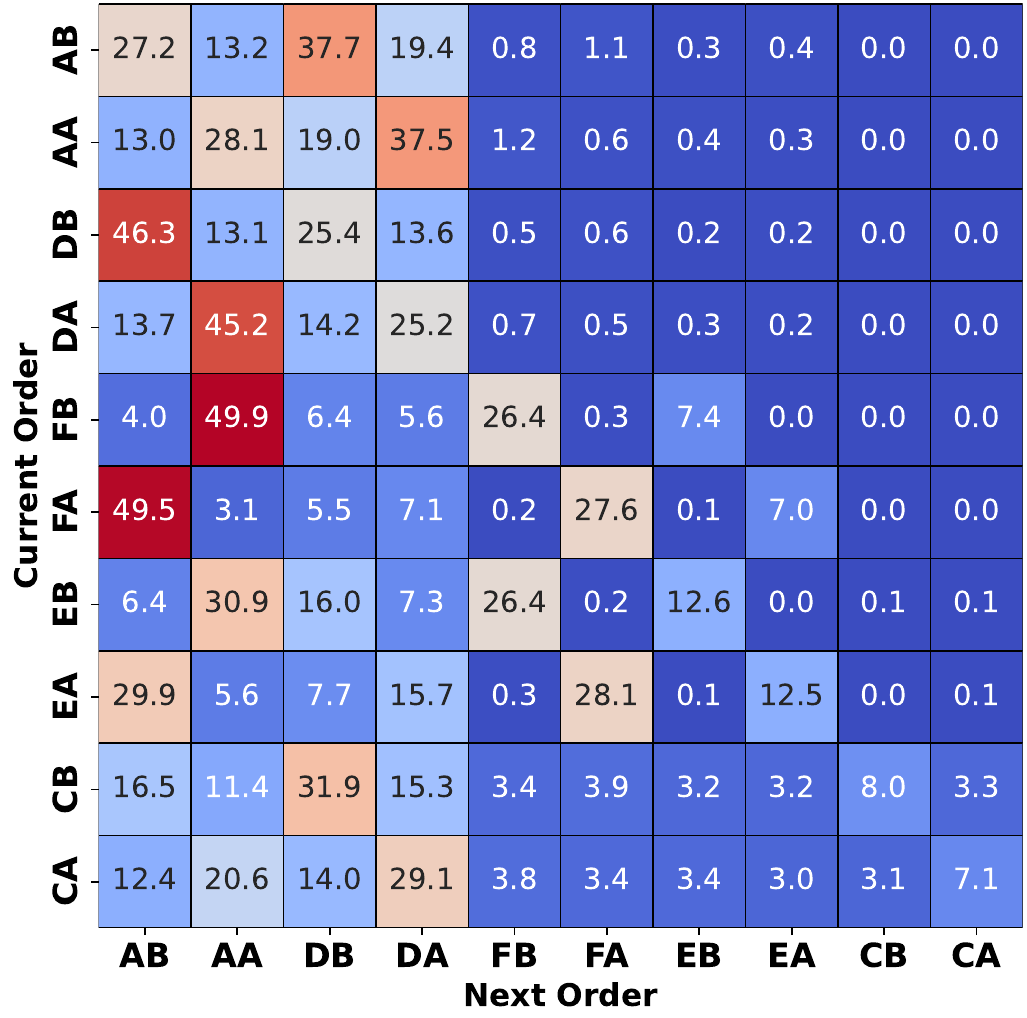}
            \caption{LMC-T4}
        \end{subfigure} \\

        \begin{subfigure}{0.3\textwidth}
            \includegraphics[width=\textwidth]{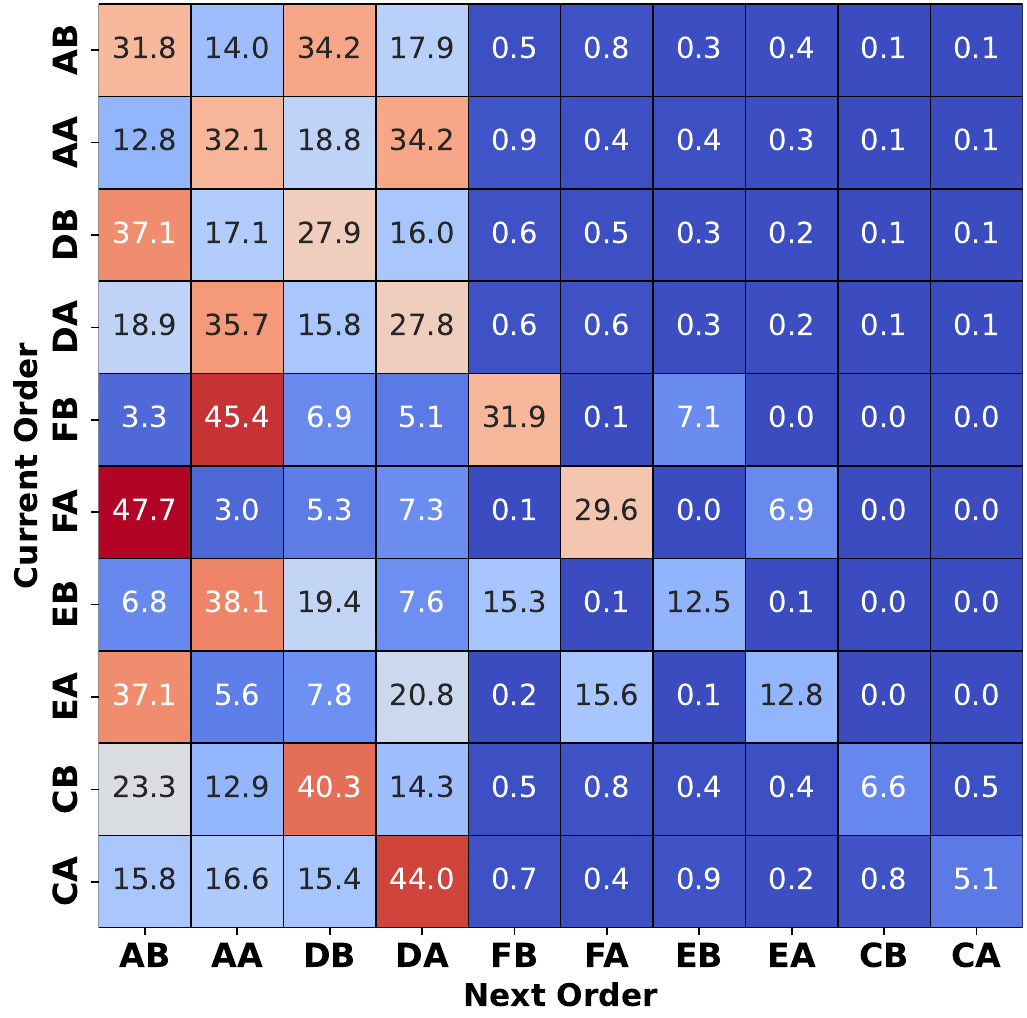}
            \caption{HMC-T5}
        \end{subfigure} &
        \begin{subfigure}{0.3\textwidth}
            \includegraphics[width=\textwidth]{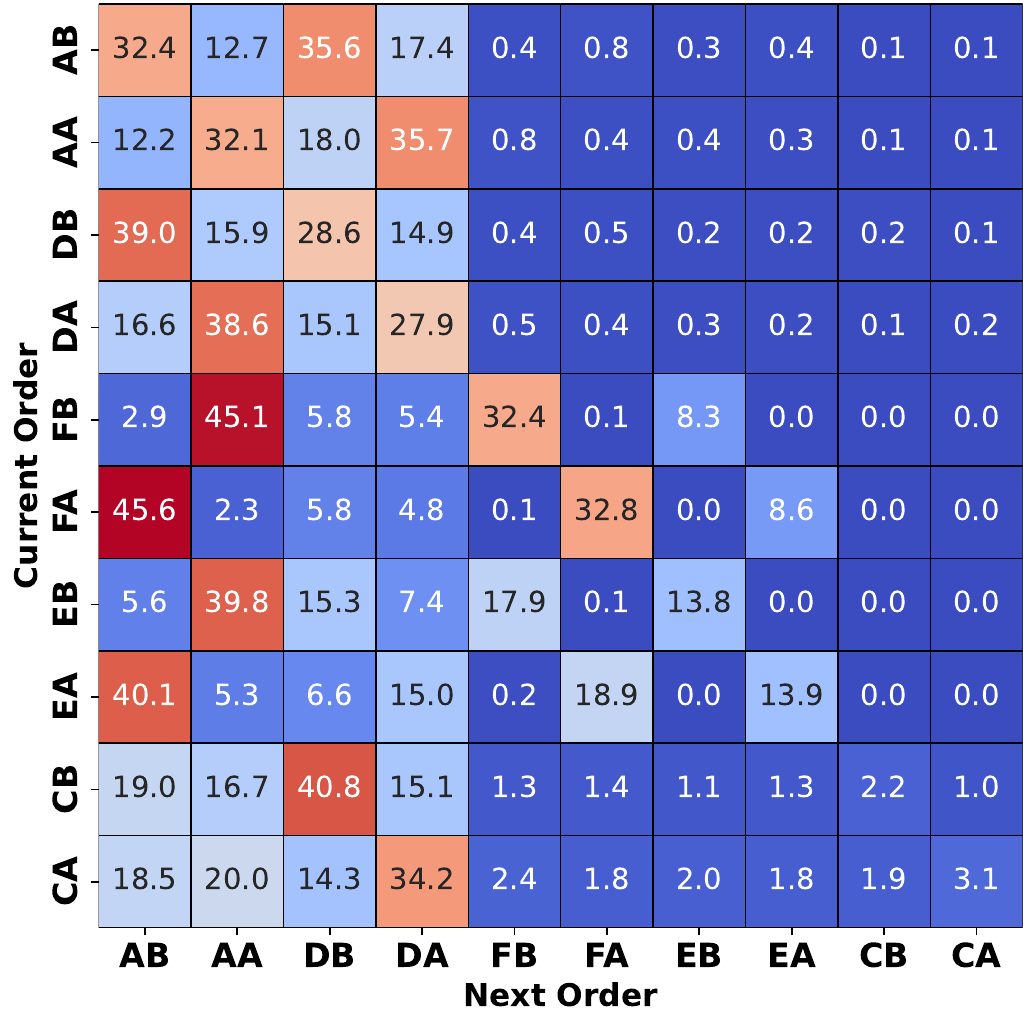}
            \caption{MMC-T5}
        \end{subfigure} &
        \begin{subfigure}{0.3\textwidth}
            \includegraphics[width=\textwidth]{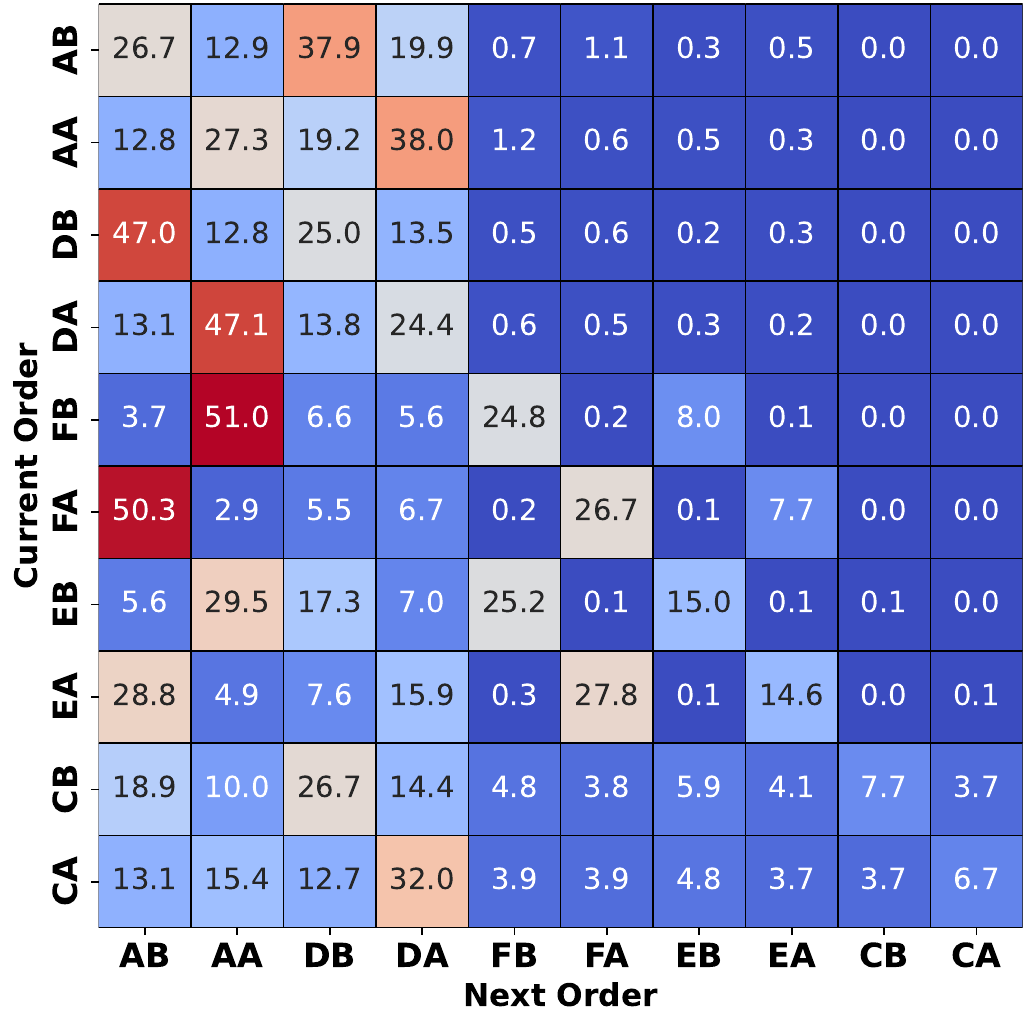}
            \caption{LMC-T5}
        \end{subfigure} \\

        \begin{subfigure}{0.3\textwidth}
            \includegraphics[width=\textwidth]{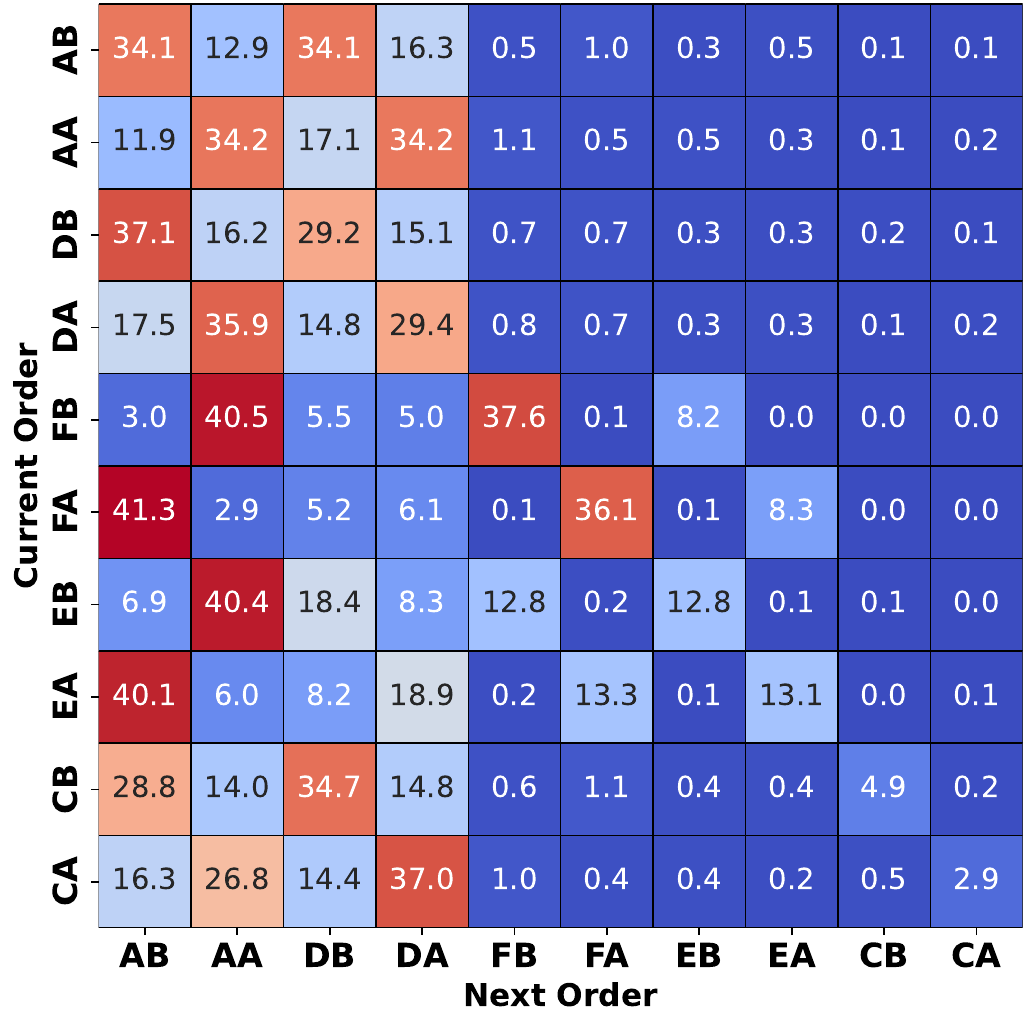}
            \caption{HMC-T6}
        \end{subfigure} &
        \begin{subfigure}{0.3\textwidth}
            \includegraphics[width=\textwidth]{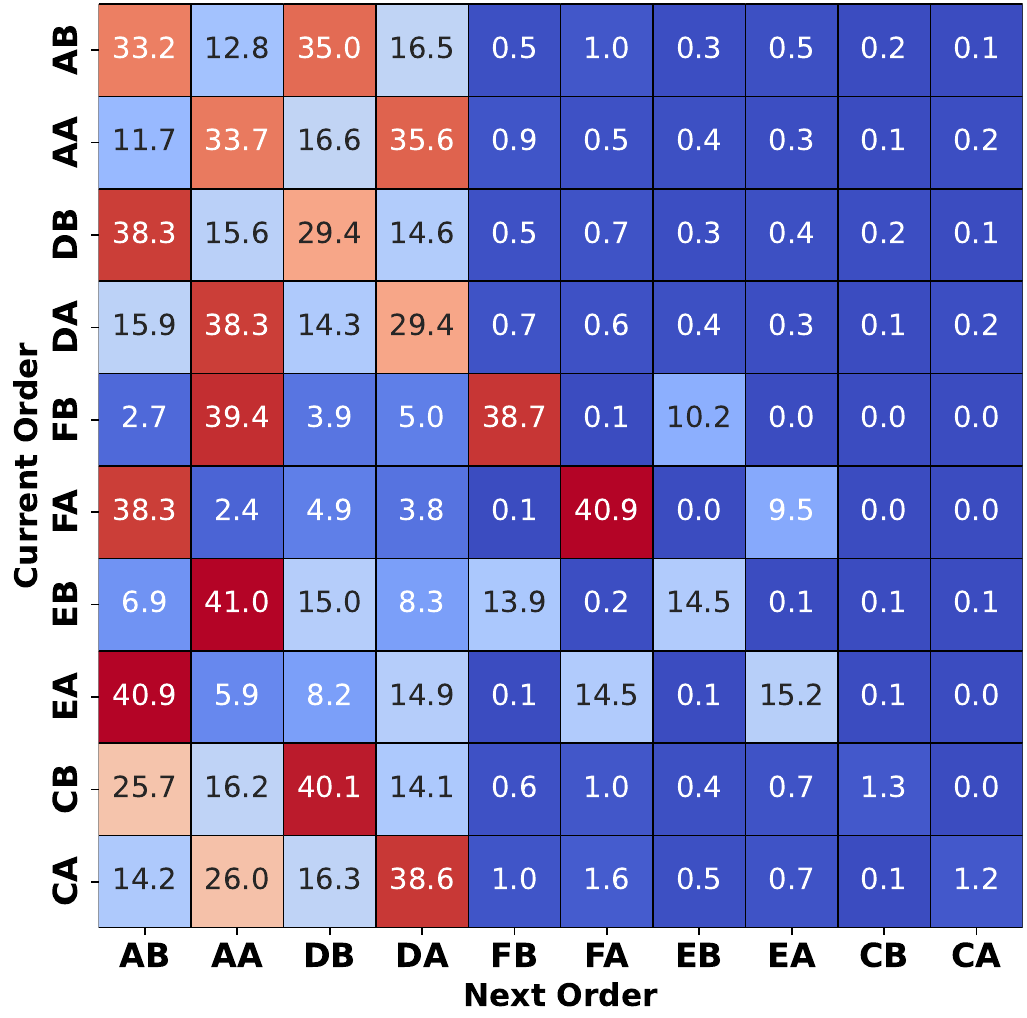}
            \caption{MMC-T6}
        \end{subfigure} &
        \begin{subfigure}{0.3\textwidth}
            \includegraphics[width=\textwidth]{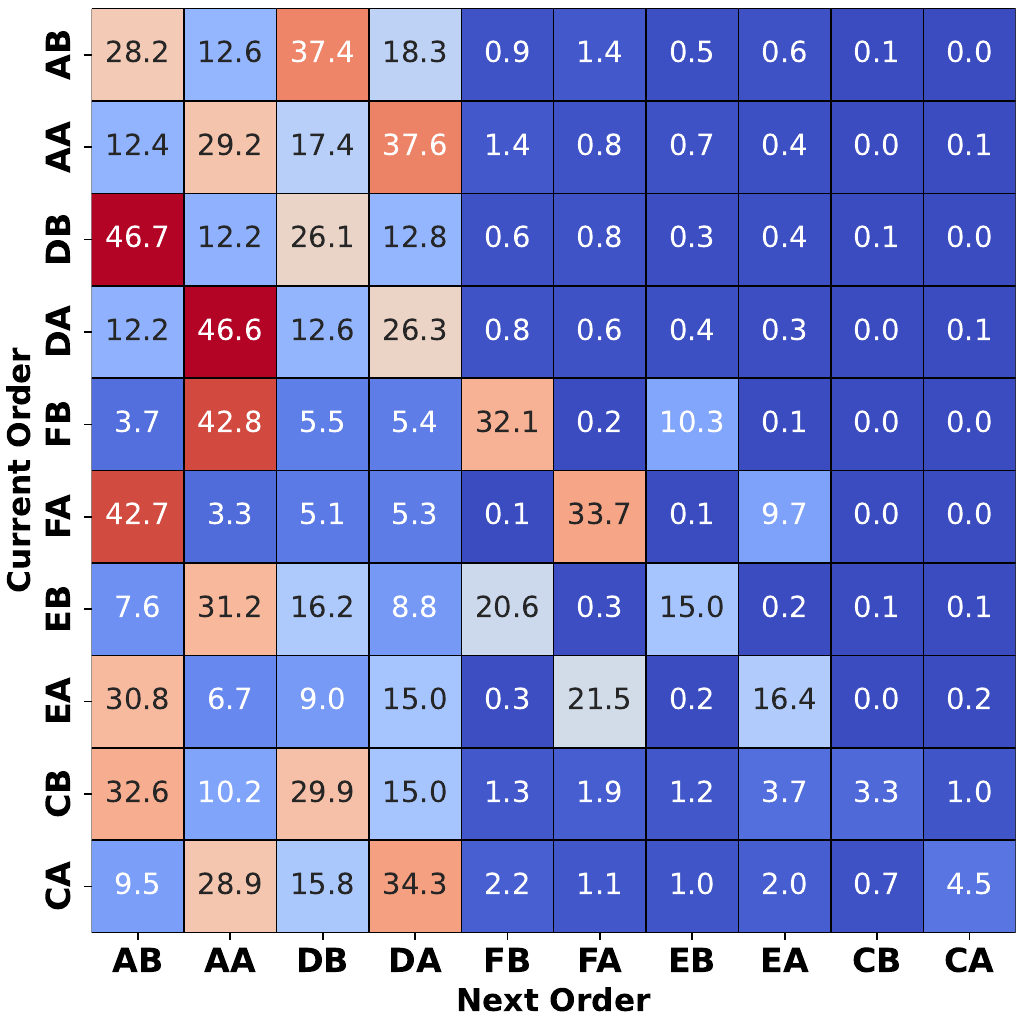}
            \caption{LMC-T6}
        \end{subfigure} \\
    \end{tabular}
    \caption{TPMs of stock market orders during different time-zones of a trading day for HMC, MMC and LMC stocks.}
    \label{fig:TPM_Order2}
\end{figure}

In the Appendix, the average count of order submissions (Fig.~\ref{fig:order_count}), average G-statistic and P-value (Table~\ref{tab:G-test}), transition probability matrices of orders (Figs.~\ref{fig:TPM_Order1} and~\ref{fig:TPM_Order2}) and stationary distribution of different orders (Table~\ref{tab:SD_Order}) at different time-zones of trading day for High Market Cap (HMC), Medium Market Cap (MMC), and Low Market Cap (LMC) stocks are presented.

\begin{table}[H]
\centering
\caption{G-Statistic and P-Value for HMC, MMC and LMC stocks during different time-zones of a trading day.}
\label{tab:G-test}
\begin{adjustbox}{width=\textwidth}
\begin{tabular}{|l|c|c|c|c|c|c|c|c|} 
\hline 
Market Cap & Stocks & \multicolumn{6}{c|}{Average G-Statistic ($\times10^{3}$)} & P-Value\\ 
\cline{3-8} 
 & & T1 & T2 & T3 & T4 & T5 & T6 &  \\ 
\hline 
\multirow{5}{*}{HMC} & AMZN & 13502.289 & 11935.074 & 14226.339 & 12665.036 & 9807.113 & 11763.720 & $\ll 0.05$ \\
\cline{2-9} 
                           & JNJ & 17.688 & 17.287 & 20.651 & 20.286 & 16.972 & 26.089 & $\ll 0.05$ \\
\cline{2-9}
                           & JPM & 5878.909 & 10508.300 & 8984.732 & 6926.353 & 7802.640 & 12020.273 & $\ll 0.05$ \\
\cline{2-9}
                           & MSFT & 21221.382 & 15368.124 & 21909.093 & 18697.975 & 18684.860 & 18568.360 & $\ll 0.05$ \\
\cline{2-9}
                           & XOM & 6712.750 & 5226.557 & 9388.656 & 3996.065 & 3942.664 & 11058.684 & $\ll 0.05$ \\
\hline
\multirow{5}{*}{MMC}   & ABBV & 9.977 & 10.642 & 12.313 & 11.477 & 10.218 & 18.482 & $\ll 0.05$ \\
\cline{2-9}
                            & HSBC & 1817.800 & 1612.320 & 13.381 & 12.402 & 9.447 & 14.481 & $\ll 0.05$ \\
\cline{2-9}
                            & NFLX & 7487.963 & 3317.286 & 2579.778 & 399.634 & 25.337 & 2868.105 & $\ll 0.05$ \\
\cline{2-9}
                            & ORCL & 12283.359 & 10499.351 & 14748.636 & 12595.168 & 7043.537 & 9334.835 & $\ll 0.05$ \\
\cline{2-9}
                            & PEP & 13.107 & 13.626 & 16.732 & 1490.033 & 13.915 & 26.232 & $\ll 0.05$ \\
\hline
\multirow{5}{*}{LMC} & AVGO & 18.058 & 16.897 & 21.099 & 18.910 & 15.420 & 29.136 & $\ll 0.05$ \\
\cline{2-9}
                             & BKNG & 8.682 & 7.403 & 9.200 & 7.009 & 6.444 & 10.895 & $\ll 0.05$ \\
\cline{2-9}
                             & BMY & 19.326 & 21.546 & 23.417 & 22.743 & 20.880 & 44.392 & $\ll 0.05$ \\
\cline{2-9}
                             & NKE & 12.371 & 13.633 & 18.152 & 17.287 & 15.698 & 28.528 & $\ll 0.05$ \\
\cline{2-9}
                             & UNP & 9.008 & 10.302 & 12.877 & 12.614 & 12.230 & 19.252 & $\ll 0.05$ \\
\hline
\end{tabular}
\end{adjustbox}
\end{table}

\begin{table}[H]
\centering
\caption{Stationary distribution of different types of orders at different time-zones for HMC, MMC and LMC stocks.}
\label{tab:SD_Order}
\begin{adjustbox}{width=\textwidth}
\begin{tabular}{|c|c|*{10}{c|}}
\hline
\multirow{2}{*}{Time-Zone} & \multirow{2}{*}{Cap} & \multicolumn{10}{c|}{Stationary Distribution} \\
\cline{3-12} 
 &  & \textbf{$\pi_{AB}$} & \textbf{$\pi_{AA}$} & \textbf{$\pi_{DB}$} & \textbf{$\pi_{DA}$} & \textbf{$\pi_{FB}$} & \textbf{$\pi_{FA}$} & \textbf{$\pi_{EB}$} & \textbf{$\pi_{EA}$} & \textbf{$\pi_{CB}$} & \textbf{$\pi_{CA}$} \\ 
\hline
\multirow{3}{*}{T1} 
 & HMC & 0.2611 & 0.2486 & 0.2335 & 0.2227 & 0.0120 & 0.0104 & 0.0049 & 0.0045 & 0.0012 & 0.0010 \\  
\cline{2-12}
 & MMC & 0.2494 & 0.2485 & 0.2313 & 0.2263 & 0.0092 & 0.0095 & 0.0039 & 0.0040 & 0.0084 & 0.0095 \\  
\cline{2-12}
 & LMC & 0.2553 & 0.2537 & 0.2342 & 0.2251 & 0.0114 & 0.0104 & 0.0046 & 0.0041 & 0.0006 & 0.0007 \\  
\hline
\multirow{3}{*}{T2} 
 & HMC & 0.2517 & 0.2454 & 0.2386 & 0.2348 & 0.0108 & 0.0087 & 0.0043 & 0.0037 & 0.0012 & 0.0009 \\  
\cline{2-12}
 & MMC & 0.2411 & 0.2489 & 0.2302 & 0.2378 & 0.0092 & 0.0091 & 0.0041 & 0.0036 & 0.0071 & 0.0089 \\  
\cline{2-12}
 & LMC & 0.2479 & 0.2482 & 0.2351 & 0.2368 & 0.0119 & 0.0104 & 0.0049 & 0.0040 & 0.0004 & 0.0005 \\  
\hline
\multirow{3}{*}{T3} 
 & HMC & 0.2526 & 0.2450 & 0.2403 & 0.2341 & 0.0100 & 0.0088 & 0.0038 & 0.0036 & 0.0010 & 0.0008 \\  
\cline{2-12}
 & MMC & 0.2513 & 0.2453 & 0.2408 & 0.2345 & 0.0091 & 0.0092 & 0.0039 & 0.0038 & 0.0011 & 0.0010 \\  
\cline{2-12}
 & LMC & 0.2523 & 0.2442 & 0.2403 & 0.2319 & 0.0112 & 0.0108 & 0.0045 & 0.0040 & 0.0004 & 0.0004 \\  
\hline
\multirow{3}{*}{T4} 
 & HMC & 0.2507 & 0.2463 & 0.2390 & 0.2365 & 0.0098 & 0.0085 & 0.0039 & 0.0035 & 0.0009 & 0.0008 \\  
\cline{2-12}
 & MMC & 0.2505 & 0.2463 & 0.2397 & 0.2363 & 0.0089 & 0.0090 & 0.0038 & 0.0037 & 0.0010 & 0.0009 \\  
\cline{2-12}
 & LMC & 0.2489 & 0.2475 & 0.2369 & 0.2348 & 0.0121 & 0.0108 & 0.0044 & 0.0039 & 0.0003 & 0.0003 \\  
\hline
\multirow{3}{*}{T5} 
 & HMC & 0.2509 & 0.2458 & 0.2387 & 0.2353 & 0.0104 & 0.0091 & 0.0041 & 0.0039 & 0.0009 & 0.0008 \\  
\cline{2-12}
 & MMC & 0.2506 & 0.2460 & 0.2403 & 0.2350 & 0.0089 & 0.0090 & 0.0041 & 0.0041 & 0.0011 & 0.0010 \\  
\cline{2-12}
 & LMC & 0.2473 & 0.2489 & 0.2356 & 0.2356 & 0.0117 & 0.0109 & 0.0048 & 0.0046 & 0.0003 & 0.0003 \\  
\hline
\multirow{3}{*}{T6} 
 & HMC & 0.2504 & 0.2463 & 0.2334 & 0.2316 & 0.0132 & 0.0121 & 0.0054 & 0.0051 & 0.0013 & 0.0012 \\  
\cline{2-12}
 & MMC & 0.2454 & 0.2497 & 0.2324 & 0.2355 & 0.0114 & 0.0124 & 0.0052 & 0.0054 & 0.0013 & 0.0012 \\  
\cline{2-12}
 & LMC & 0.2444 & 0.2501 & 0.2270 & 0.2323 & 0.0154 & 0.0150 & 0.0073 & 0.0068 & 0.0009 & 0.0009 \\  
\hline
\end{tabular}
\end{adjustbox}
\end{table}

\end{document}